\documentclass[aps, superscriptaddress, showpacs]{revtex4-1}
\pdfoutput=1

\usepackage{graphicx}
\usepackage{amssymb,amsmath}
\usepackage{bm}
\usepackage{color}

\begin{document}

\title{Optimizing mutual synchronization of rhythmic spatiotemporal patterns in reaction-diffusion systems}

\author{Yoji Kawamura}
\email{ykawamura@jamstec.go.jp}
\affiliation{Department of Mathematical Science and Advanced Technology,
Japan Agency for Marine-Earth Science and Technology, Yokohama 236-0001, Japan}

\author{Sho Shirasaka}
\email{shirasaka@neuron.t.u-tokyo.ac.jp}
\affiliation{Research Center for Advanced Science and Technology, University of Tokyo, Tokyo 153-8904, Japan}

\author{Tatsuo Yanagita}
\email{yanagita@isc.osakac.ac.jp}
\affiliation{Osaka Electro-Communication University, Neyagawa 572-8530, Japan}

\author{Hiroya Nakao}
\email{nakao@mei.titech.ac.jp (corresponding author)}
\affiliation{Department of Systems and Control Engineering,
Tokyo Institute of Technology, Tokyo 152-8552, Japan}
\affiliation{Department of Mechanical Engineering, University of California, Santa Barbara, California}

\begin{abstract}
Optimization of the stability of synchronized states between a pair of symmetrically coupled reaction-diffusion systems exhibiting rhythmic spatiotemporal patterns is studied in the framework of the phase reduction theory.
The optimal linear filter that maximizes the linear stability of the in-phase synchronized state is derived for the case where the two systems are nonlocally coupled.
The optimal nonlinear interaction function that theoretically gives the largest linear stability of the in-phase synchronized state is also derived.
The theory is illustrated by using typical rhythmic patterns in FitzHugh-Nagumo systems as examples.
\end{abstract}


\maketitle

\section{Introduction}

Synchronization of rhythmic systems is widely observed in the real world and has been extensively studied in various areas of science and engineering~\cite{ref:winfree80,Kuramoto,Glass,ref:pikovsky01,ref:hoppensteadt97,ref:ermentrout10,bullo}. In biological systems, synchronization often plays significant functional roles, such as generation of heartbeats, rhythmic gaits, and circadian rhythms. In some engineering systems, synchronization is a precondition for their functionality, where power grids provide a well-known example.

Recently, synchronization between non-conventional self-oscillatory systems has recently attracted considerable attention, for example, dynamical systems with time-delayed feedback~\cite{Kotani,Novicenko}, spatially extended reaction-diffusion systems~\cite{Fukushima,Hildebrand,Nakao}, and fluid systems~\cite{read2010,read2012,Kawamura2}.
In analyzing synchronization properties of rhythmic systems described as weakly perturbed limit-cycle oscillators, the phase reduction theory has been used as a standard method for clarifying their mechanisms
~\cite{Kuramoto,ref:winfree80,Glass,ref:pikovsky01,ref:hoppensteadt97,ref:ermentrout10,ref:brown04,ref:nakao16,ref:ashwin16}.
Recent developments in the phase reduction theory~\cite{Kotani,Novicenko,Nakao,Kawamura2} have shown that, even if the state space of the rhythmic system is infinite-dimensional, it can still exhibit synchronization in a similar way to low-dimensional oscillators
as long as it exhibits stable limit-cycle oscillation.

In this study, we consider synchronization of reaction-diffusion systems exhibiting rhythmic spatiotemporal patterns.
Reaction-diffusion systems have played important roles in modeling a variety of spatiotemporal patterns that arise in chemical and biological systems~\cite{ref:mikhailov06,ref:mikhailov13,ref:mikhailov14,ref:cross93,ref:cross09,ref:engel14,ref:engel16}.
Among them, rhythmic spatiotemporal patterns, such as oscillating spots, target waves, and rotating spirals, can be regarded as stable limit-cycle oscillations of the reaction-diffusion systems.
Synchronization between rhythmic spatiotemporal patterns has been experimentally realized using
coupled electrochemical systems exhibiting reaction waves of ${\rm H}_2 {\rm O}_2$ reduction on ${\rm Pt}$ ring electrodes, where two waves are coupled via the common chemical solution~\cite{Fukushima}, and
coupled photosensitive Belousov-Zhabotinsky systems 
exhibiting spiral patterns, where the two patterns are coupled via video cameras and projectors~\cite{Hildebrand} (see also~\cite{Sakurai,ref:mikhailov06,Epstein,ref:showalter15,ref:weiss15,ref:weiss17}).
Synchronization of rhythmic fluid flows has also been studied and its possible importance in global climate has been argued~\cite{read2010,read2012}.

In our recent work~\cite{Nakao}, we generalized the conventional phase reduction theory for finite-dimensional limit-cycle oscillators to limit-cycle oscillations of reaction-diffusion systems with infinite-dimensional state space.
Using the theory, we derived the phase sensitivity function, which characterizes linear phase response of the rhythmic pattern to weak perturbations, and analyzed mutual synchronization between a pair of reaction-diffusion systems coupled by linear diffusive interaction.
We also developed similar phase reduction theories for the collective oscillations in globally coupled noisy oscillators~\cite{Kawamura1} and for oscillatory thermal convection in a Hele-Shaw cell~\cite{Kawamura2}.
Moreover, we analyzed synchronization between non-interacting convection cells exhibiting oscillatory thermal convection caused by common noise and derived the optimal input pattern for stable noise-induced synchronization~\cite{Kawamura3}.

In this study, we consider the case that a pair of reaction-diffusion systems, both of which are exhibiting rhythmic spatiotemporal patterns, are mutually coupled via weak symmetric interaction.
In the case of the simplest linear diffusive interaction where every point of the system is coupled to a corresponding point of the other system, we have shown that the two systems can undergo mutual synchronization~\cite{Nakao}.
However, the phase sensitivity function of the rhythmic pattern is often strongly localized in space, and introducing interaction at every point in the system as in Ref.~\cite{Nakao} would not be generally efficient.
In this study, we aim to clarify the theoretical limit of efficiency in synchronizing rhythmic patterns via mutual coupling by seeking for optimal interaction schemes that realize stable in-phase synchronization between two reaction-diffusion systems in the framework of the phase reduction theory.

Regarding optimization of synchronization, optimal input signals that efficiently entrains a limit-cycle oscillator described by ordinary differential equations have been obtained for various situations~\cite{Zlotnik,ref:moehlis06,ref:harada10,ref:dasanayake11,ref:zlotnik12,ref:zlotnik12,ref:pikovsky15-prl,ref:tanaka14a,ref:tanaka14b,ref:tanaka15,ref:zlotnik2016}.
Also, in our preceding article, we derived optimal cross-coupling matrices that maximize linear stability of the synchronized states in a pair of diffusively coupled limit-cycle oscillators described by ordinary differential equations~\cite{Shirasaka}.
In this paper, we further generalize the analysis to a pair of coupled reaction-diffusion systems exhibiting rhythmic patterns and try to derive the optimal interaction function.
We first restrict ourselves to a practical situation where the interaction between the two systems is linear and derive the optimal filtering function for stable synchronization.
We then derive the optimal nonlinear interaction function between the systems to clarify the theoretical limit to the improvement of stability.
The results are illustrated by using rhythmic spatiotemporal patterns in FitzHugh-Nagumo reaction-diffusion systems, that is, a traveling pulse and an oscillating spot in one dimension, and a rotating spiral in two dimensions.

\section{Theory}

\subsection{A pair of mutually coupled reaction-diffusion systems}

We consider a pair of mutually coupled $m$-component reaction-diffusion systems in a $d$-dimensional space exhibiting stable limit-cycle oscillations (see Figures for typical rhythmic spatiotemporal patterns of the FitzHugh-Nagumo reaction-diffusion systems), described by
\begin{align}
  \frac{\partial}{\partial t}{\bm X}_1({\bm r}, t)
  &= {\bm F}({\bm X}_1, {\bm r}) + \hat{D} \nabla^2 {\bm X}_1({\bm r}, t)
  + \epsilon \int_V d\bm{r}' \, \hat{A}\left( \bm{r}, \bm{r}' \right) \bm{X}_2\left( \bm{r}', t \right),
  \cr
  \frac{\partial}{\partial t}{\bm X}_2({\bm r}, t)
  &= {\bm F}({\bm X}_2, {\bm r}) + \hat{D} \nabla^2 {\bm X}_2({\bm r}, t)
  + \epsilon \int_V d\bm{r}' \, \hat{A}\left( \bm{r}, \bm{r}' \right) \bm{X}_1\left( \bm{r}', t \right),
  \label{linearRD}
\end{align}
in some spatial domain $V \subset {\bm R}^d$.
Here, ${\bm r} \in {\bm R}^d$ is the spatial location, $t \in {\bm R}$ is the time, ${\bm X}_{1, 2} : {\bm R}^d \times {\bm R} \to {\bm R}^m$ are the spatial patterns of the systems, i.e., the system states, ${\bm F} : {\bm R}^m \times {\bm R}^d \to {\bm R}^m$ represents the dynamics of the system, $\hat{D} \in  {\bm R}^{m \times m}$ is a matrix of diffusion constants, and $\nabla^2$ is the Laplacian operator.
For simplicity, we consider two identical systems whose dynamics are described by the same function ${\bm F}$ and diffusion matrix $\hat{D}$, and also assume that they are symmetrically coupled.
The dynamics ${\bm F}$ can depend on the location ${\bm r}$, for example, the excitability of the system can be different from place to place.
The last term in the right-hand side of each equation represents mutual linear interaction between the two systems, where each system is coupled to the other system via a $m \times m$ matrix of spatial linear filters ${\hat A}({\bm r}, {\bm r}') : {\bm R}^d \times {\bm R}^d \to {\bm R}^{m \times m}$.
The parameter $\epsilon \geq 0$ is the interaction intensity, which is assumed to be small.

As a benchmark, we also consider the following systems with simple, direct mutual interaction:
\begin{align}
  \frac{\partial}{\partial t}{\bm X}_1({\bm r}, t)
  &= {\bm F}({\bm X}_1, {\bm r}) + \hat{D} \nabla^2 {\bm X}_1({\bm r}, t)
  + \epsilon \bm{X}_2\left( \bm{r}, t \right),
  \cr
  \frac{\partial}{\partial t}{\bm X}_2({\bm r}, t)
  &= {\bm F}({\bm X}_2, {\bm r}) + \hat{D} \nabla^2 {\bm X}_2({\bm r}, t)
  + \epsilon \bm{X}_1\left( \bm{r}, t \right),
  \label{directRD}
\end{align}
where every point in the system is directly coupled to the corresponding point of the other system without filtering. 
The definitions of the variables and parameters are the same as in Eq.~(\ref{linearRD}).
Both interaction schemes, Eq.~(\ref{linearRD}) and Eq.~(\ref{directRD}), can exhibit in-phase synchronization between the systems, and we compare the stability of the synchronized states between them.

As another, typical interaction scheme, we may also consider diffusive interaction between the two reaction-diffusion systems given by
\begin{align}
  \frac{\partial}{\partial t}{\bm X}_1({\bm r}, t)
  &= {\bm F}({\bm X}_1, {\bm r}) + \hat{D} \nabla^2 {\bm X}_1({\bm r}, t)
  + \epsilon \int_V d\bm{r}' \, \hat{A}\left( \bm{r}, \bm{r}' \right)
  \left[ \bm{X}_2\left( \bm{r}', t \right) - \bm{X}_1\left( \bm{r}', t \right) \right],
  \cr
  \frac{\partial}{\partial t}{\bm X}_2({\bm r}, t)
  &= {\bm F}({\bm X}_2, {\bm r}) + \hat{D} \nabla^2 {\bm X}_2({\bm r}, t)
  + \epsilon \int_V d\bm{r}' \, \hat{A}\left( \bm{r}, \bm{r}' \right)
  \left[ \bm{X}_1\left( \bm{r}', t \right) - \bm{X}_2\left( \bm{r}', t \right) \right].
\end{align}
It is clear that this interaction scheme also allows in-phase synchronization.
For sufficiently small $\epsilon$, which is assumed throughout this study, we can show that the stability of the in-phase synchronized state with this diffusive interaction scheme is approximately equal (up to $O(\epsilon)$) to that for the interaction scheme given by Eq.~(\ref{linearRD}).
Similarly, in our previous paper~\cite{Nakao}, we analyzed the following simple case with direct diffusive interaction, where every point in the system is diffusively coupled to the corresponding point in the other system as
\begin{align}
  \frac{\partial}{\partial t}{\bm X}_1({\bm r}, t)
  &= {\bm F}({\bm X}_1, {\bm r}) + \hat{D} \nabla^2 {\bm X}_1({\bm r}, t)
  + \epsilon \left[ \bm{X}_2\left( \bm{r}, t \right) - \bm{X}_1\left( \bm{r}, t \right) \right],
  \cr
  \frac{\partial}{\partial t}{\bm X}_2({\bm r}, t)
  &= {\bm F}({\bm X}_2, {\bm r}) + \hat{D} \nabla^2 {\bm X}_2({\bm r}, t)
  + \epsilon \left[ \bm{X}_1\left( \bm{r}, t \right) - \bm{X}_2\left( \bm{r}, t \right) \right],
\end{align}
and showed that the two systems undergo mutual synchronization by using the phase reduction theory.
Linear stability of the in-phase synchronized state of this interaction scheme is also approximately the same as that for Eq.~(\ref{directRD}) when $\epsilon$ is sufficiently small.

In this study, we focus on the interaction schemes given by Eq.~(\ref{linearRD}) and Eq.~(\ref{directRD}) and analyze their synchronization properties.
We consider a general nonlocal interaction given by Eq.~(\ref{linearRD}) and try to optimize the linear filter $\hat{A}({\bm r}, {\bm r}')$ so that the two systems exhibit more stable in-phase synchronization than the case with the simple interaction given by Eq.~(\ref{directRD}).
In the following, we refer to the interaction scheme in Eq.~(\ref{linearRD}) as {\em nonlocal}, while that in Eq.~(\ref{directRD}) as {\em direct}.

\subsection{Phase reduction}

In Ref.~\cite{Nakao}, we generalized the phase reduction theory for finite-dimensional limit-cycle oscillators~\cite{Kuramoto} to reaction-diffusion systems exhibiting stable rhythmic patterns.
Using the theory, we can systematically approximate the dynamics of two weakly coupled reaction-diffusion systems by simple two-dimensional coupled phase equations and analyze their synchronization properties.

Suppose that a single reaction-diffusion system (without interaction) exhibits a stable rhythmic pattern, that is, a stable limit-cycle solution ${\bm X}_0({\bm r}, t+T) = {\bm X}_0({\bm r}, t)$ of period $T$ and frequency $\omega = 2\pi / T$.
We can introduce a phase variable $\theta \in [0, 2\pi]$ of the rhythmic pattern around the limit-cycle solution in the state space of the system that always increases with a constant frequency $\omega$ in the absence of perturbation (i.e., the {\it asymptotic phase}~\cite{ref:winfree80,Kuramoto,ref:hoppensteadt97,ref:ermentrout10,ref:brown04,ref:nakao16,ref:ashwin16}), and represent the limit-cycle solution as a function of phase $\theta$, rather than time $t$, as ${\bm X}_0({\bm r}, \theta)$ ($0 \leq \theta \leq 2\pi$).

In the phase reduction theory, the phase sensitivity function ${\bm Z}({\bm r}, \theta) : {\bm R}^d \times [0, 2\pi] \to {\bm R}^m$ plays an important role, which characterizes linear phase response of the system to a weak perturbation that is applied when the system state is at ${\bm X}_0({\bm r}, \theta)$.
It is given by a $2\pi$-periodic solution (an eigenfunction associated with the zero eigenvalue) to the adjoint equation
\begin{align}
  \frac{\partial}{\partial \theta} {\bm Z}({\bm r}, \theta) =
  - D{\bm F}({\bm X}_0({\bm r}, \theta), {\bm r})^{\dag} {\bm Z}({\bm r}, \theta)
  - \hat{D}^{\dag} \nabla^2 {\bm Z}({\bm r}, \theta)
\end{align}
with appropriate boundary conditions, where $D{\bm F}$ is the Jacobi matrix of ${\bm F}({\bm X}, {\bm r})$ at ${\bm X} = {\bm X}_0({\bm r}, \theta)$ and $\dag$ denotes matrix transpose, and satisfies a normalization condition
\begin{align}
  \int_V d{\bm r} \, {\bm Z}({\bm r}, \theta) \cdot {\bm U}({\bm r}, \theta) = 1
  \label{normalization}
\end{align}
for $0 \leq \theta \leq 2\pi$.
Here, we defined the tangent field ${\bm U}({\bm r}, \theta)$ of ${\bm X}_0({\bm r}, \theta)$ along the limit-cycle solution as ${\bm U}({\bm r}, \theta) = {\partial {\bm X}_0({\bm r}, \theta)} / {\partial \theta}$.

We consider weakly coupled reaction-diffusion systems given by Eq.~(\ref{linearRD}) or Eq.~(\ref{directRD}), and assume that the rhythmic patterns are only slightly perturbed and persist even when weak mutual interaction between the two systems is introduced.
We can then approximately describe the system states using only scalar phase variables $\theta_{1, 2} \in [0, 2\pi]$ as ${\bm X}_{1,2}({\bm r}, t) = {\bm X}_0({\bm r}, \theta_{1,2}(t))$, and derive approximate phase equations for $\theta_{1, 2}(t)$ from Eq.~(\ref{linearRD}) or Eq.~(\ref{directRD}) as
\begin{align}
  \dot{\theta}_1(t) &= \omega + \epsilon \Gamma(\theta_1 - \theta_2),
  \cr
  \dot{\theta}_2(t) &= \omega + \epsilon \Gamma(\theta_2 - \theta_1),
  \label{phase0}
\end{align}
where the overdot ($\dot{}$) represents $d/dt$ and the $2\pi$-periodic function $\Gamma(\phi) : [0, 2\pi] \to {\bm R}$ is called the phase coupling function.
In the case of the nonlocal interaction, Eq.~(\ref{linearRD}), the phase coupling function is given by
\begin{align}
  \Gamma(\phi)
  = \frac{1}{2\pi} \int_{0}^{2\pi} d\psi \int_V d\bm{r} \int_V d\bm{r}' \,
  \bm{Z}\left( \bm{r}, \psi + \phi \right) \cdot \hat{A}\left( \bm{r}, \bm{r}' \right) \bm{X}_0\left( \bm{r}', \psi \right),
  \label{phasecpl0}
\end{align}
and in the case of the direct interaction, Eq.~(\ref{directRD}), the phase coupling function is simply given by
\begin{align}
  \Gamma(\phi)
  = \frac{1}{2\pi} \int_{0}^{2\pi} d\psi \int_V d\bm{r} \,
  \bm{Z}\left( \bm{r}, \psi + \phi \right) \cdot \bm{X}_0\left( \bm{r}, \psi \right).
\end{align}

Synchronization between the two systems can be analyzed in the same way as for finite-dimensional coupled oscillators~\cite{Kuramoto}.
From Eq.~(\ref{phase0}), we can derive the dynamics of the phase difference $\phi = \theta_1 - \theta_2$, which we restrict in the range $[-\pi, \pi]$, as
\begin{align}
  \dot{\phi}(t) = \epsilon \Gamma_a(\phi),
  \label{phasedifference}
\end{align}
where
\begin{align}
  \Gamma_a(\phi) = \Gamma(\phi) - \Gamma(-\phi)
\end{align}
is the antisymmetric part of the phase coupling function $\Gamma(\phi)$.
Fixed points of this equation satisfying $\Gamma_a(\phi^*) = 0$ correspond to the phase differences where the two systems exhibit synchronization.
The stability of a fixed point $\phi = \phi_*$ is characterized by the slope of $\Gamma_a(\phi_*)$.

Because the function $\Gamma_a(\phi)$ vanishes at $\phi = 0$ and $\phi = \pm \pi$ by definition, Eq.~(\ref{phasedifference}) has fixed points at $\phi^*=0$ and $\phi^* = \pm \pi$.
We focus on the in-phase synchronized state, $\phi^* = 0$, whose stability is characterized by
\begin{align}
  \Gamma_a'(0) = \left. \frac{d}{d\phi} \Gamma_a(\phi) \right|_{\phi = 0}.
\end{align}
In the case of the nonlocal interaction, Eq.~(\ref{linearRD}), this value depends on ${\hat A}({\bm r}, {\bm r}')$.
For the direct interaction, Eq.~(\ref{directRD}), this value can be explicitly calculated as
\begin{align}
  \Gamma_a'(0) = 2 \Gamma'(0) 
  &= 2 \cdot \frac{1}{2\pi} \int_{0}^{2\pi} d\psi \int_V d\bm{r} \,
  \frac{\partial \bm{Z}\left( \bm{r}, \psi \right)}{\partial \psi} \cdot
  \bm{X}_0\left( \bm{r}, \psi \right)
  \cr
  &= - 2 \cdot \frac{1}{2\pi} \int_{0}^{2\pi} d\psi \int_V d\bm{r} \,
  \bm{Z}\left( \bm{r}, \psi \right) \cdot
  {\bm U}\left( \bm{r}, \psi \right) = -2,
\end{align}
where we performed partial integration, used the $2\pi$-periodicity of ${\bm Z}$ and ${\bm X}_0$ to eliminate the surface terms, and used the normalization condition Eq.~(\ref{normalization}) for ${\bm Z}$.
In this study, we try to make $\Gamma_a'(0)$ as negative as possible in order to improve the stability of the in-phase synchronization under a constraint for the norm of ${\hat A}({\bm r}, {\bm r}')$.

\subsection{Optimal linear filter for stable synchronization}

Under the framework of the phase reduction approximation for sufficiently small $\epsilon$, we seek for the optimal filter function ${\hat A}({\bm r}, {\bm r}')$ that maximizes the linear stability
$-\Gamma'_a(0)$ of the in-phase synchronized state.
As a constraint, we fix the spatial average of the Frobenius norm (or the Hilbert-Schmidt norm) of the linear filter $A({\bm r}, {\bm r}')$, i.e.,
\begin{align}
  \int_V d{\bm r} \int_V d{\bm r}' \, \left\| \hat{A}({\bm r}, {\bm r}') \right\|_F^2 = P,
\end{align}
where $P > 0$ is a constant and 
\begin{align}
  \left\| \hat{A}({\bm r}, {\bm r}') \right\|_F^2 = \sum_{i=1}^m \sum_{j=1}^m A_{ij}({\bm r}, {\bm r}')^2.
\end{align}
To this end, we consider the action
\begin{align}
  S\{\hat{A}, \lambda\} = - \Gamma_a'(0)
  - \lambda \left( \int_V d{\bm r} \int_V d{\bm r}' \sum_{i=1}^m \sum_{j=1}^m A_{ij}\left( \bm{r}, \bm{r}' \right)^2 - P \right),
\end{align}
where $\lambda$ is the Lagrange multiplier, and find $\hat{A}$ and $\lambda$ that give the extremum of $S$.

From Eq.~(\ref{phasecpl0}), the slope $\Gamma'(\phi) = d\Gamma(\phi) / d\phi$ of the phase coupling function is given by
\begin{align}
  \Gamma'(\phi)
  = \frac{1}{2\pi} \int_{0}^{2\pi} d\psi \int_V d\bm{r} \int_V d\bm{r}' \,
  \frac{\partial \bm{Z}\left( \bm{r}, \psi + \phi \right) }{\partial\psi} \cdot
  \hat{A}\left( \bm{r}, \bm{r}' \right) \bm{X}_0\left( \bm{r}', \psi \right).
\end{align}
Thus, the slope $\Gamma_a'(0)$ of the antisymmetric part $\Gamma_a(\phi) = \Gamma(\phi) - \Gamma(-\phi)$ at $\phi=0$ is calculated as
\begin{align}
  \Gamma_a'(0) = 2 \Gamma'(0)
  &= 2 \cdot \frac{1}{2\pi} \int_{0}^{2\pi} d\psi \int_V d\bm{r} \int_V d\bm{r}' \,
  \frac{\partial \bm{Z}\left( \bm{r}, \psi \right)}{\partial \psi} \cdot
  \hat{A}\left( \bm{r}, \bm{r}' \right) \bm{X}_0\left( \bm{r}', \psi \right)
  \cr
  &= - 2 \cdot \frac{1}{2\pi} \int_{0}^{2\pi} d\psi \int_V d\bm{r} \int_V d\bm{r}' \,
  \bm{Z}\left( \bm{r}, \psi \right) \cdot
  \hat{A}\left( \bm{r}, \bm{r}' \right) {\bm U}\left( \bm{r}', \psi \right).
\end{align}
Denoting the vector components of ${\bm Z}$, ${\bm X}_0$, and ${\bm U} = \partial {\bm X_0} / \partial \psi$ as $Z_1, \cdots, Z_m$, $X_1, \cdots, X_m$, and $U_1, \cdots, U_m$, respectively, and the matrix components of $\hat{A}$ as $\{ A_{ij} \}$ ($i, j = 1, 2, \cdots, m$), $\Gamma_a'(0)$ can be expressed as
\begin{align}
  \Gamma_a'(0)
  &= -2 \cdot \frac{1}{2\pi} \int_{0}^{2\pi} d\psi \int_V d\bm{r} \int_V d\bm{r}' \sum_{i=1}^m \sum_{j=1}^m
  Z_i\left( \bm{r}, \psi \right) A_{ij}\left( \bm{r}, \bm{r}' \right) U_j\left( \bm{r}', \psi \right)
  \cr
  &= -2 \, \int_V d\bm{r} \int_V d\bm{r}' \sum_{i=1}^m \sum_{j=1}^m
  A_{ij}\left( \bm{r}, \bm{r}' \right) W_{ij}\left( \bm{r}, \bm{r}' \right).
\end{align}
Here, we have defined a correlation matrix $\hat{W}\left( \bm{r}, \bm{r}' \right)$ whose components are given by
\begin{align}
  W_{ij}\left( \bm{r}, \bm{r}' \right)
  = \frac{1}{2\pi} \int_{0}^{2\pi} d\psi \,
  Z_i\left( \bm{r}, \psi \right) U_j\left( \bm{r}', \psi \right),
  \label{correlationmat}
\end{align}
which characterizes the spatial correlation between ${\bm Z}$ and ${\bm U}$ averaged over one period of oscillation.

The action is now given by
\begin{align}
  S\{ \hat{A}, \lambda \}
  &= 2 \, \int_V d\bm{r} \int_V d\bm{r}' \sum_{i=1}^m \sum_{j=1}^m
  A_{ij}\left( \bm{r}, \bm{r}' \right) W_{ij}\left( \bm{r}, \bm{r}' \right)
  \cr
  &- \lambda \left( \int_V d{\bm r} \int_V d{\bm r}' \sum_{i=1}^m \sum_{j=1}^m A_{ij}\left( \bm{r}, \bm{r}' \right)^2 - P \right),
\end{align}
and by taking variations with respect to $A_{ij}({\bm r}, {\bm r}')$, we obtain
\begin{align}
  W_{ij}\left( \bm{r}, \bm{r}' \right) = \lambda A_{ij}\left( \bm{r}, \bm{r}' \right),
\end{align}
so the optimal linear filter is given by
\begin{align}
  {\hat A}({\bm r}, {\bm r}') = \frac{1}{\lambda} {\hat W}({\bm r}, {\bm r}').
\end{align}
Differentiating $S$ with respect to $\lambda$ simply gives the constraint,
\begin{align}
  \int_V d{\bm r} \int_V d{\bm r}' \sum_{i=1}^m \sum_{j=1}^m A_{ij}\left( \bm{r}, \bm{r}' \right)^2 = P.
\end{align}
This equation gives the Lagrange multiplier $\lambda$ as
\begin{align}
  \lambda
  = \sqrt{ \frac{1}{P} \int_V d{\bm r} \int_V d{\bm r}' \sum_{i=1}^m \sum_{j=1}^m W_{ij}\left( \bm{r}, \bm{r}' \right)^2 },
\end{align}
where we have chosen the plus sign so that $S$ takes the extremum (or $\Gamma'(0)$ becomes negative) at the optimal $\hat{A}$.
The largest negative slope of $\Gamma_a'(0)$ is given by
\begin{align}
  \Gamma_a'(0)
  = - \frac{2}{\lambda} \int_V d{\bm r} \int_V d{\bm r}' \sum_{i=1}^m \sum_{j=1}^m W_{ij}\left( \bm{r}, \bm{r}' \right)^2.
\end{align}
This value gives the largest stability of the in-phase synchronized state for the nonlocal interaction.

Note that spatial linear filtering of the field ${\bm X}$ by $\hat{A}$ in Eq.~(\ref{linearRD}) is a fundamental method of image processing and can be easily performed.
It is also notable that the expression for the correlation matrix, Eq.~(\ref{correlationmat}), consists of only two terms, ${\bm U}$ and ${\bm Z}$. These are the most fundamental quantities of limit-cycling systems, that is, ${\bm U}$ is a tangent field to the limit-cycle orbit and ${\bm Z}$ is a phase sensitivity function of a limit-cycle orbit, which are adjoint to each other. These quantities always arise in the phase reduction analysis of coupled oscillators and the expression of Eq.~(\ref{correlationmat}) is physically natural.

\subsection{Optimal nonlinear interaction}

Theoretically, we can consider a more general case that the two system are coupled via nonlinear functional as
\begin{align}
  \frac{\partial}{\partial t}{\bm X}_1({\bm r}, t)
  &= {\bm F}({\bm X}_1, {\bm r}) + \hat{D} \nabla^2 {\bm X}_1({\bm r}, t)
  + \epsilon {\bm H} \{{\bm X}_2(\cdot, t), {\bm r}\},
  \cr
  \frac{\partial}{\partial t}{\bm X}_2({\bm r}, t)
  &= {\bm F}({\bm X}_2, {\bm r}) + \hat{D} \nabla^2 {\bm X}_2({\bm r}, t)
  + \epsilon {\bm H} \{{\bm X}_1(\cdot, t), {\bm r}\},
  \label{nonlinearRD}
\end{align}
where ${\bm H} : C \times {\bm R}^d \to {\bm R}^m$ ($C$ represents the set of spatial patterns) is a functional of the spatial pattern.
The phase coupling function $\Gamma(\phi) : [0, 2\pi] \to {\bm R}$ in this case can be calculated as
\begin{align}
  \Gamma(\phi)
  = \frac{1}{2\pi} \int_0^{2\pi} d\psi \int_V d{\bm r} \, {\bm Z}({\bm r}, \phi + \psi) \cdot {\bm H}({\bm r}, \psi),
\end{align}
where ${\bm H}({\bm r}, \psi) = {\bm H}\{{\bm X}_0(\cdot, \psi), {\bm r}\}$.
Note that we have replaced the functional ${\bm H}\{{\bm X}_0(\cdot, \psi), {\bm r}\}$ of ${\bm X}_0({\bm r}, \psi)$ by the function ${\bm H}({\bm r}, \psi)$ of $\psi$ here, because the function ${\bm X}_0({\bm r}, \psi)$ is solely determined by phase $\psi$ in the framework of the phase reduction.
We refer to ${\bm H}({\bm r}, \psi)$ as the interaction function.

We try to make $\Gamma_a'(0)$ as negative as possible under the constraint
\begin{align}
  \frac{1}{2\pi} \int_0^{2\pi} d\psi \int_V d{\bm r} \, \left\| {\bm H}({\bm r}, \psi) \right\|^2 = Q,
  \label{constraint}
\end{align}
that is, we fix the squared mean of ${\bm H}$ over the space and all possible combinations of the phase variables at some constant $Q>0$.
This amounts to fixing the average ``energy'' of the mutual interaction between the two systems.
We consider the following action functional of ${\bm H}$ and a Lagrange multiplier $\lambda$:
\begin{align}
  S\{ {\bm H}, \lambda \} =
  - \Gamma_a'(0)
  - \lambda \left( \frac{1}{2\pi} \int_0^{2\pi} d\psi \int_V d{\bm r} \, \left\| {\bm H}({\bm r}, \psi) \right\|^2 - Q \right),
\end{align}
where $\Gamma_a'(0)$ in the first term can be represented as
\begin{align}
  \Gamma_a'(0) = 2 \Gamma'(0)
  = 2 \cdot \frac{1}{2\pi} \int_0^{2\pi} d\psi \int_V d{\bm r} \,
  \left( \frac{\partial}{\partial \psi} {\bm Z}({\bm r}, \psi) \right) \cdot {\bm H}({\bm r}, \psi).
\end{align}
By taking variations of $S\{{\bm H}, \lambda\}$ with respect to ${\bm H}$, we obtain a Euler-Lagrange equation
\begin{align}
  - \frac{\partial}{\partial \psi}{\bm Z}({\bm r}, \psi) - \lambda {\bm H}({\bm r}, \psi) = 0,
\end{align}
which yields
\begin{align}
  {\bm H}({\bm r}, \psi) = -\frac{1}{\lambda} \frac{\partial}{\partial \psi}{\bm Z}({\bm r}, \psi).
\end{align}
The Lagrange multiplier $\lambda$ is given by
\begin{align}
  \lambda = \sqrt{
    \frac{1}{Q} \frac{1}{2\pi} \int_0^{2\pi} d\psi \int_V d{\bm r} \,
    \left\| \frac{\partial}{\partial \psi}{\bm Z}({\bm r}, \psi) \right\|^2 }.
\end{align}
Note that the plus sign, which gives the extremum of $S\{\bm H, \lambda\}$, has been chosen here.
The largest negative slope of the antisymmetric part of the phase coupling function at $\phi = 0$ is given by
\begin{align}
  \Gamma_a'(0)
  = - \frac{2}{\lambda} \frac{1}{2\pi} \int_0^{2\pi} d\psi \int_V d{\bm r} \,
  \left\| \frac{\partial}{\partial \psi} {\bm Z}({\bm r}, \psi) \right\|^2,
\end{align}
which yields the largest possible stability of the in-phase synchronized state for general nonlinear interaction of Eq.~(\ref{nonlinearRD}).

Thus, ${\bm H}({\bm r}, \psi) \propto - \partial {\bm Z}({\bm r}, \psi)/\partial \psi$ is the optimal interaction function in the nonlinear case.
That is, the optimal nonlinear interaction between the two systems is realized by (i) measuring the phase $\psi$ of the other system and (ii) driving the system using the negative derivative of the phase sensitivity function with respect to phase $\psi$.
This result is consistent with Zlotnik {\it et al.}'s~\cite{Zlotnik} for the optimal periodic input signal that maximizes linear stability of the entrainment of ordinary limit-cycle oscillators.

In practice, however, online continuous-time estimation of the instantaneous phase value $\psi$ from an observed rhythmic pattern is generally not straightforward (note that we need to estimate the correct {\it asymptotic} phase of a given pattern).
For low-dimensional oscillators, we could construct a mapping from the oscillator states to phase values beforehand and use it to estimate the instantaneous phase value of the observed oscillator state in real time.
However, construction of such a mapping from high-dimensional data of spatial patterns to phase values is generally a non-trivial, demanding task.
Thus, the optimal nonlinear interaction may not be easy to realize experimentally, in contrast to the the nonlocal interaction that can easily be implemented once the optimal linear filter $\hat{A}$ is given.

\section{Numerical simulations}

\subsection{FitzHugh-Nagumo model}

We now illustrate the theoretical results with numerical examples.
As the reaction-diffusion system, we use the FitzHugh-Nagumo (FHN) system described by two field variables, ${\bm X} = (X_u, X_v) = (u, v)$, which obey
\begin{align}
  \frac{\partial}{\partial t} u({\bm r}, t) &= u ( u - \alpha )(1 - u) - v + D_u \nabla^2 u,
  \cr
  \frac{\partial}{\partial t} v({\bm r}, t) &= \tau^{-1} ( u - \gamma v ) + D_v \nabla^2 v.
\end{align}
This system can exhibit various rhythmic spatiotemporal patterns such as traveling pulses, oscillating spots, target waves, and rotating spirals.
Here, as typical examples, we consider a traveling pulse (on a ring) and an oscillating spot in one-dimensional systems, and a rotating spiral in two-dimensional systems.
The setup of simulations are basically the same as in Ref.~\cite{Nakao}, but some of the parameter values and system sizes are modified. 

We compare the results for the nonlocal interaction, Eq.~(\ref{linearRD}) with the optimal linear filter $\hat{A}$, the direct interaction, Eq.~(\ref{directRD}), and the optimal nonlinear interaction, Eq.~(\ref{nonlinearRD}) with the optimal function ${\bm H}$.
To make a fair comparison between different interaction schemes, we fix the squared mean of the interaction term over the spatial domain and over the phase to a constant, namely, we calculate the quantity
\begin{align}
  I = \frac{1}{2\pi} \int_0^{2\pi} d\psi \int_V d{\bm r} \, \left\| {\bm X}_0({\bm r}, \psi) \right\|^2
\end{align}
for the direct interaction and appropriately normalize the nonlocal interaction,
\begin{align}
  {\bm G}({\bm r}, \theta) = \int_V d{\bm r}' \, \hat{A}({\bm r}, {\bm r}') {\bm X}_0({\bm r}', \theta)
  = \frac{1}{\lambda} \int_V d{\bm r}' \, \hat{W}({\bm r}, {\bm r}') {\bm X}_0({\bm r}', \theta),
\end{align}
and the nonlinear interaction,
\begin{align}
  {\bm H}({\bm r}, \theta) = -\frac{1}{\lambda} \frac{\partial}{\partial \theta}{\bm Z}({\bm r}, \theta),
\end{align}
so that
\begin{align}
  \frac{1}{2\pi} \int_0^{2\pi} d\psi \int_V d{\bm r} \, \left\| {\bm G}({\bm r}, \psi) \right\|^2 = I
\end{align}
and
\begin{align}
  \frac{1}{2\pi} \int_0^{2\pi} d\psi \int_V d{\bm r} \, \left\| {\bm H}({\bm r}, \psi) \right\|^2 = I
\end{align}
are satisfied.
That is, we fix the average ``energy'' of the interaction functions between the two systems over one period of oscillation.

For each pattern, the limit-cycle solution, phase sensitivity function, and antisymmetric part of the phase coupling functions for the direct, optimal nonlocal, and optimal nonlinear interactions are shown.
Direct numerical simulations of the synchronization process of the reaction-diffusion systems are also shown for the direct and optimal nonlocal interactions (see Appendix on numerical computation of the optimal nonlocal interaction).
Optimal nonlinear interaction is not simulated, because it requires the instantaneous phase values of the spatial patterns and therefore is not easy to realize.
We use the synchronization error, $E = \sqrt{\int_V d{\bm r} \, \| {\bm X}_1({\bm r}, t) - {\bm X}_2({\bm r}, t) \|^2}$, and the phase difference, $\phi = \theta_1 - \theta_2$, measured from the simulated patterns to see the convergence of the systems to synchronization.
Here, the phase difference is measured stroboscopically by using threshold-crossing times of the patterns roughly at intervals of the period of oscillation.

\subsection{Traveling pulse}

\begin{figure}
\centering
\includegraphics[width=0.65\hsize]{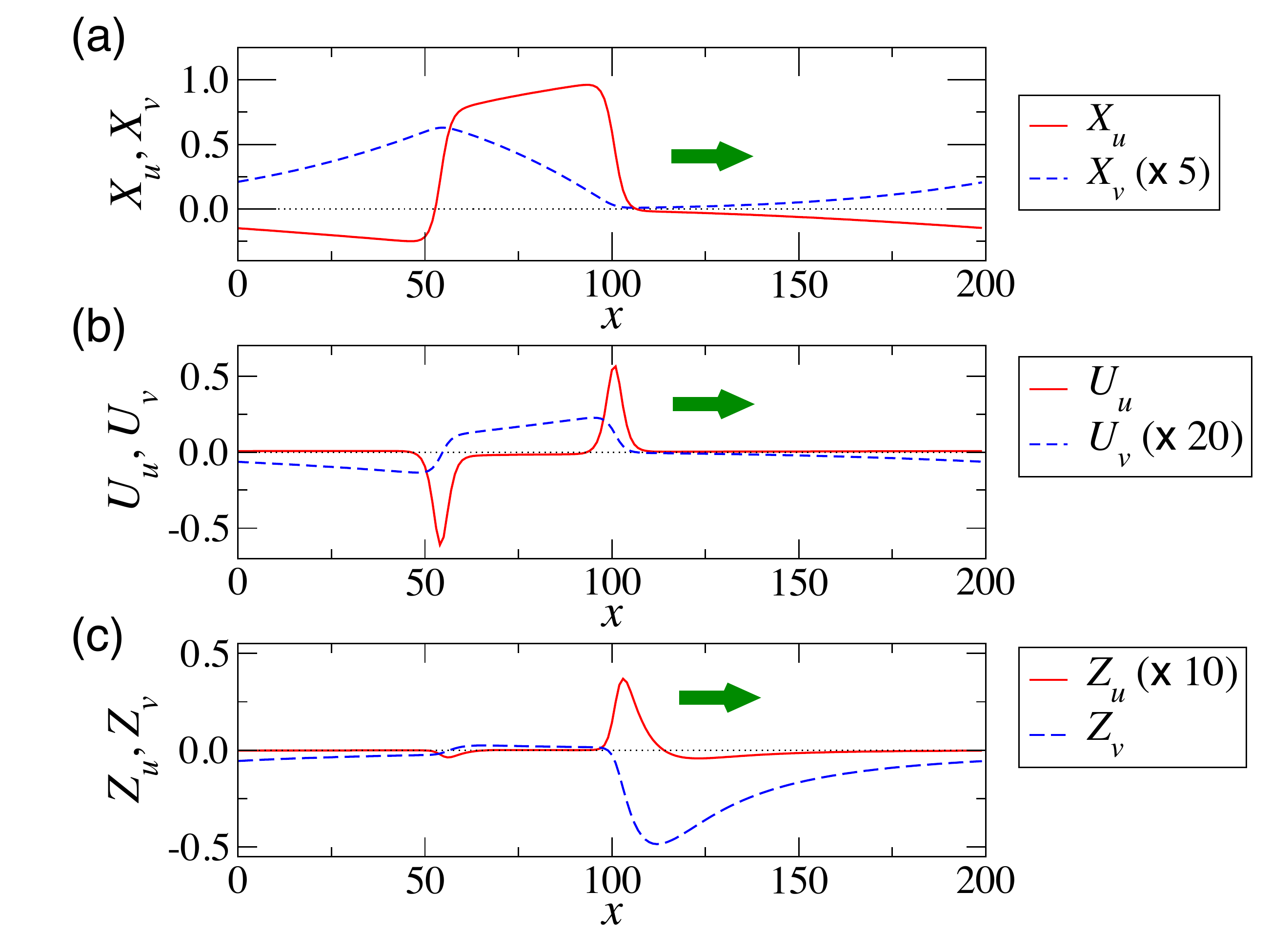}
\caption{Traveling-pulse solution of the FHN model.
  Snapshots of (a) ${\bm X}_0(x, \theta=0) = (X_u, X_v)$, (b) ${\bm U}(x, \theta=0) = (U_u, U_v)$, and (c) ${\bm Z}(x, \theta=0) = (Z_u, Z_v)$.
  Both $u$ and $v$ components are plotted in each figure. Some of the curves are enlarged for visibility.}
\label{FigA0}
\end{figure}

\begin{figure}
\centering
\includegraphics[width=0.55\hsize]{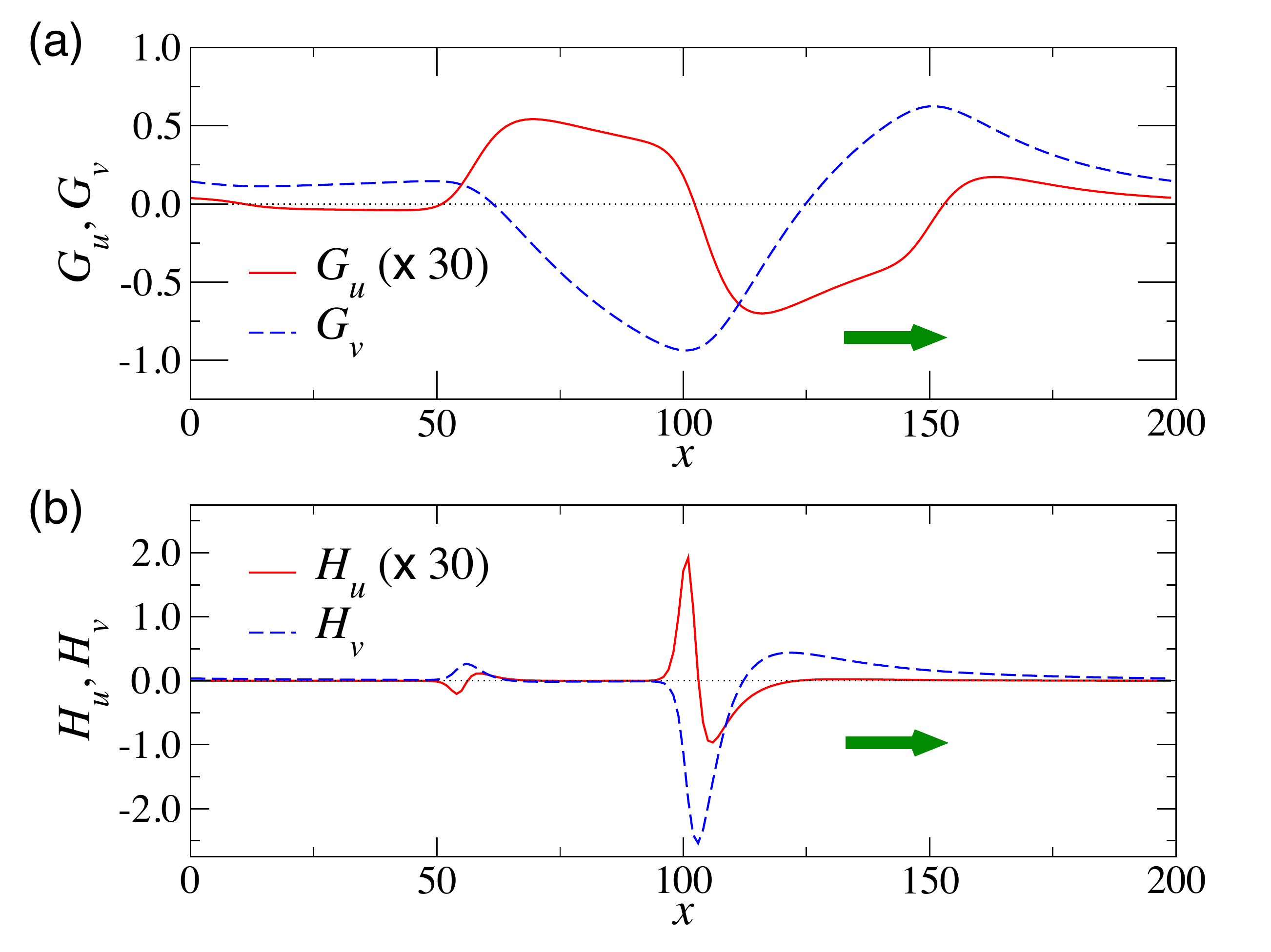}
\caption{Optimized interaction functions for the traveling pulse.
  (a) Optimal nonlocal interaction ${\bm G}(x, \theta=0) = (G_u, G_v)$ and (b) optimal nonlinear interaction ${\bm H}(x, \theta=0) = (H_u, H_v)$.
  Both $u$ and $v$ components are plotted. Results for the $u$ components are enlarged for clarity.}
\label{FigA1}
\end{figure}

\begin{figure}
\centering
\includegraphics[width=0.55\hsize]{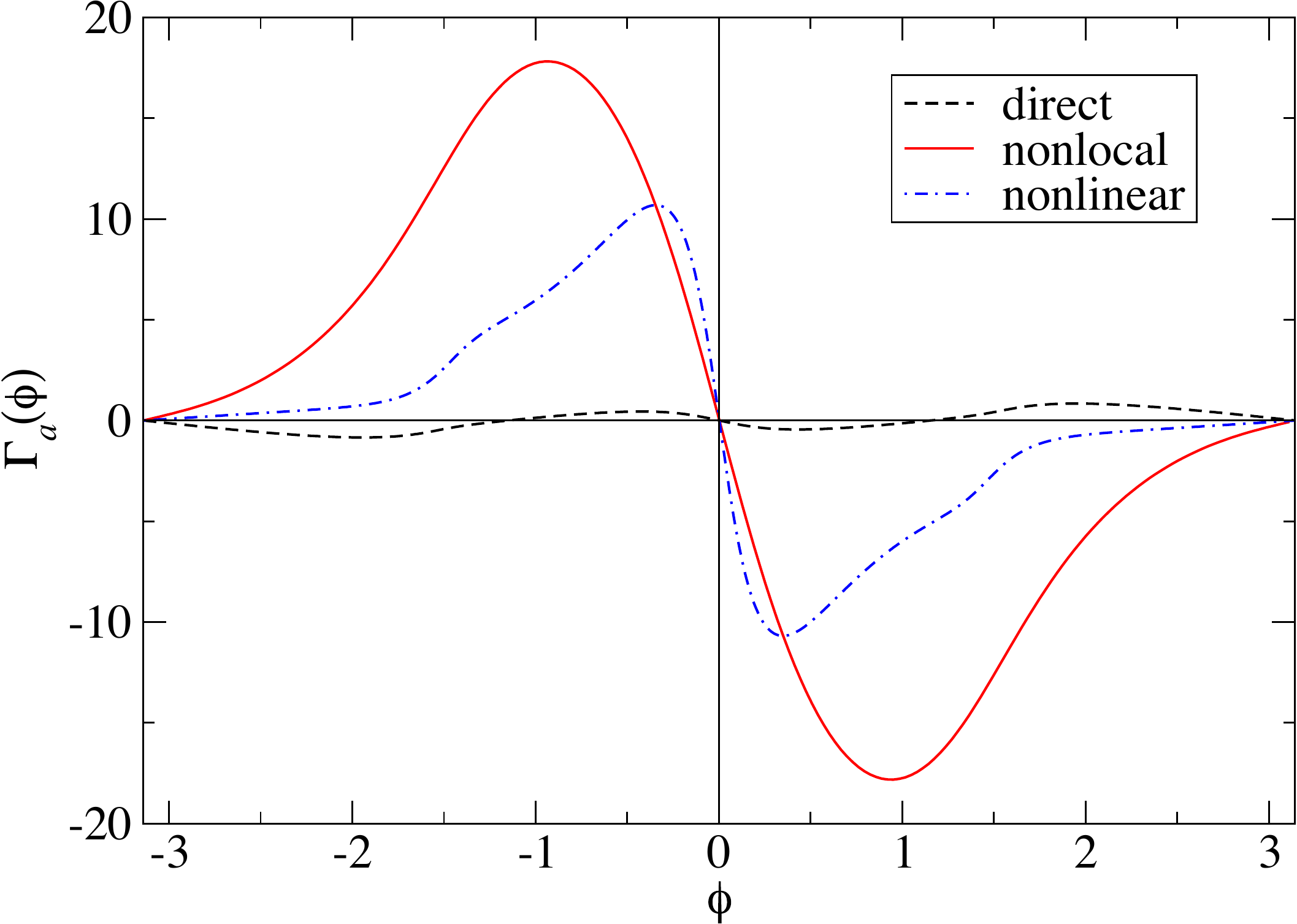}
\caption{Antisymmetric part $\Gamma_a(\phi)$ of the phase coupling functions for the traveling pulse.
  Results for direct interaction, optimal nonlocal interaction, and optimal nonlinear interaction are shown.}
\label{FigA2}
\end{figure}

\begin{figure}
\centering
\includegraphics[width=0.45\hsize]{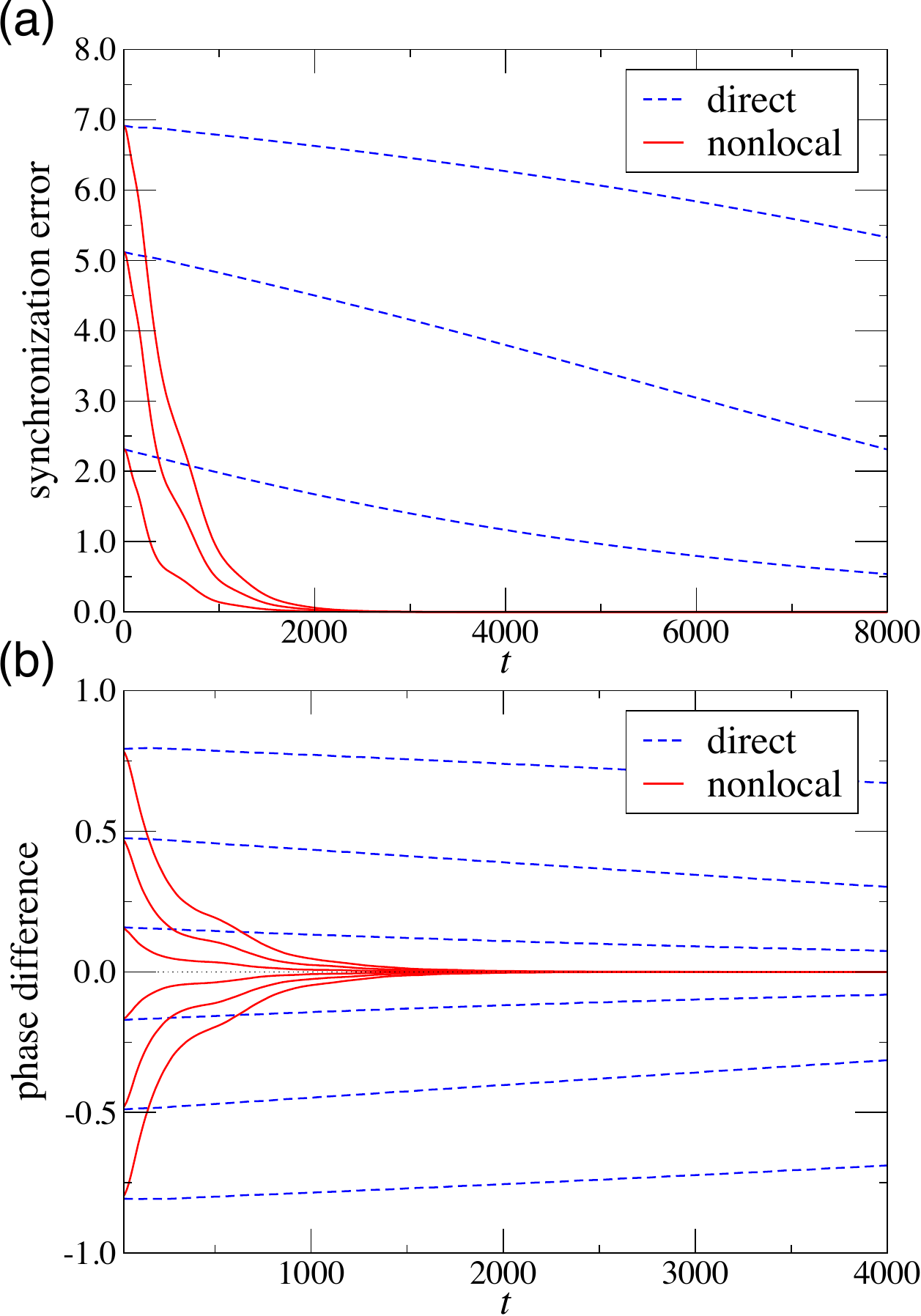}
\caption{Synchronization dynamics between traveling pulses.
  Results for direct interaction and for optimal nonlocal interaction are compared.
  (a) Evolution of synchronization error. (b) Evolution of phase difference.}
\label{FigA3}
\end{figure}

Figures~\ref{FigA0}-\ref{FigA3} show the results for a traveling pulse in a one-dimensional space $0 \leq x \leq L$ with periodic boundary conditions.
The parameters are $\alpha = 0.1$, $\tau^{-1} = 0.002$, $\gamma = 2.5$, $D_u = 1.0$, and $D_v = 0.1$.
The system size is $L=200$ and discretized by using $N=200$ grid points.
The interaction intensity between the systems is $\epsilon = 0.0001$.

Figure~\ref{FigA0} shows the traveling-pulse solution ${\bm X}_0(x, \theta)$,
the tangent function ${\bm U}(x, \theta) = \partial {\bm X}_0(x, \theta) / \partial \theta$,
and the phase sensitivity function ${\bm Z}(x, \theta)$.
In each figure, both $u$ and $v$ components of ${\bm X}_0(x,\theta=0) = (X_u, X_v)$, ${\bm U}(x, \theta=0) = (U_u, U_v)$, and ${\bm Z}(x, \theta=0) = (Z_u, Z_v)$ are plotted.
Because the pulse keeps a constant shape and simply travels to the right with a constant velocity, all the functions simply translate to the right with a constant velocity without changing their shapes.
The period of the oscillation is $T \approx 395.7$ and the frequency is $\omega \approx 0.01588$.
From these data, the correlation matrix $\hat{W}(x, x')$ is calculated.

Figure~\ref{FigA1} shows the nonlocal interaction ${\bm G}(x, \theta)$ and the nonlinear interaction ${\bm H}(x, \theta)$ at $\theta = 0$.
These optimal interaction functions also translate to the right with the pulse without changing their shapes.
The nonlocal interaction function and nonlinear interaction function differ from each other.
However, they have one thing in common; that is, both interaction functions change their signs in front of the pulse.
This is actually essential for efficient control of the phase of the pulse.

Figure~\ref{FigA2} shows the antisymmetric part $\Gamma_a(\phi)$ of the phase coupling function for the direct, nonlocal optimal, and optimal nonlinear interaction.
Both nonlocal interaction and nonlinear interaction yield much higher stability of the synchronized state than the direct interaction.
The nonlinear interaction gives the highest linear stability, but the nonlocal interaction also yields reasonably high stability.
We can also observe that both nonlocal interaction and nonlinear interaction uniquely give the in-phase synchronized state as the globally stable solution, while the direct interaction yields both in-phase and anti-phase synchronized solutions as stable solutions.
Because global stability of the solutions is not considered in the present optimization, these results are coincidental.

Figure~\ref{FigA3} shows the synchronization dynamics between the two reaction-diffusion systems obtained by direct numerical simulations for the direct and optimal nonlocal interactions.
Temporal evolution of the synchronization error and the phase difference (measured stroboscopically at each period of oscillation) are shown.
We can clearly see that the optimal nonlocal interaction yields much faster synchronization than the direct coupling.
When the phase difference is sufficiently small, exponential growth (or decay) rate of the phase difference coincides with the slope $\Gamma_a'(0)$.

\subsection{Oscillating spot}

\begin{figure}
\centering
\includegraphics[width=0.65\hsize]{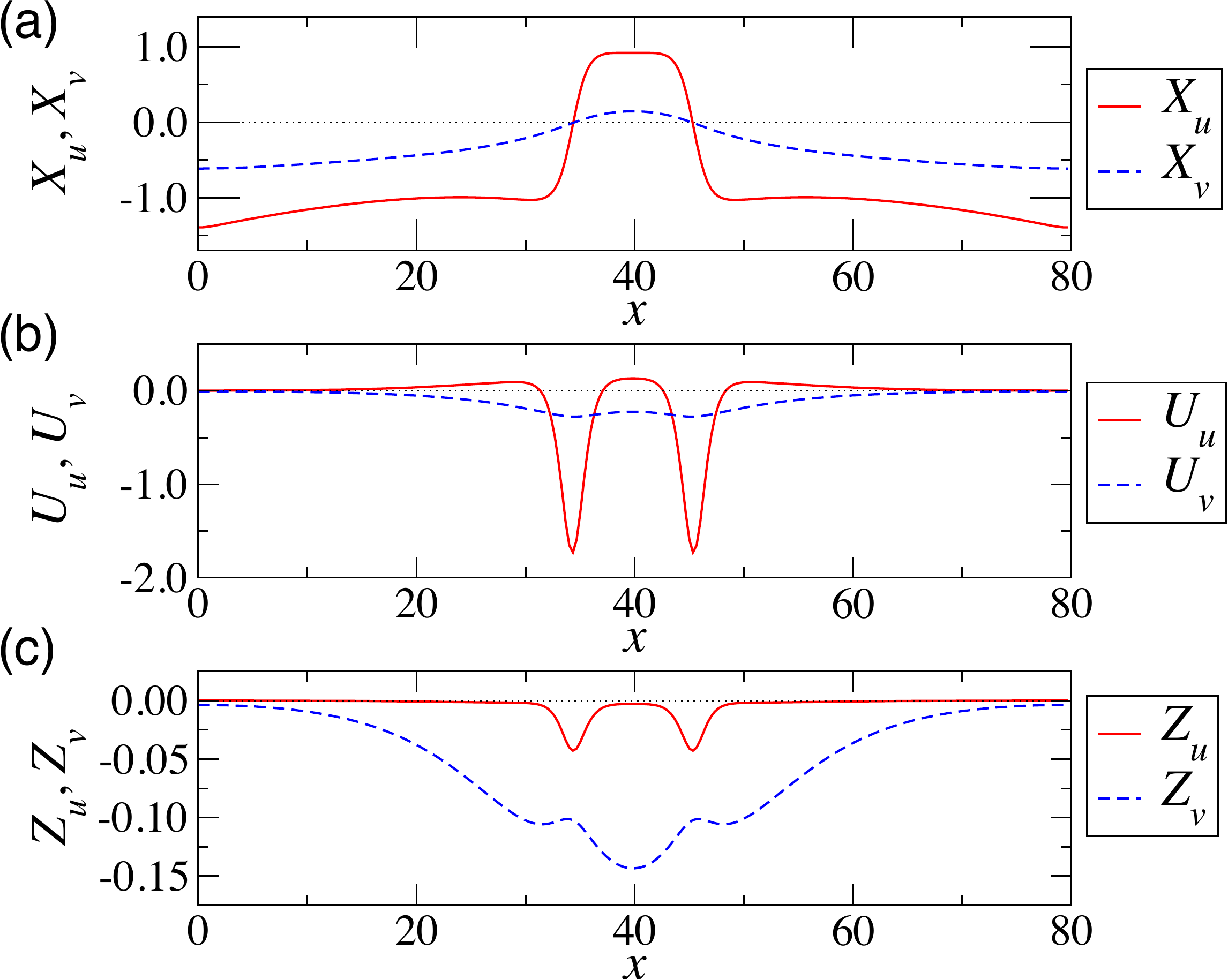}
\caption{Oscillating-spot solution of the FHN model.
  Snapshots of (a) ${\bm X}_0(x, \theta=0) = (X_u, X_v)$, (b) ${\bm U}(x, \theta=0) = (U_u, U_v)$, and (c) ${\bm Z}(x, \theta=0) = (Z_u, Z_v)$.
  Both $u$ and $v$ components are shown.}
\label{FigB0}
\end{figure}

\begin{figure}
\centering
\includegraphics[width=0.6\hsize]{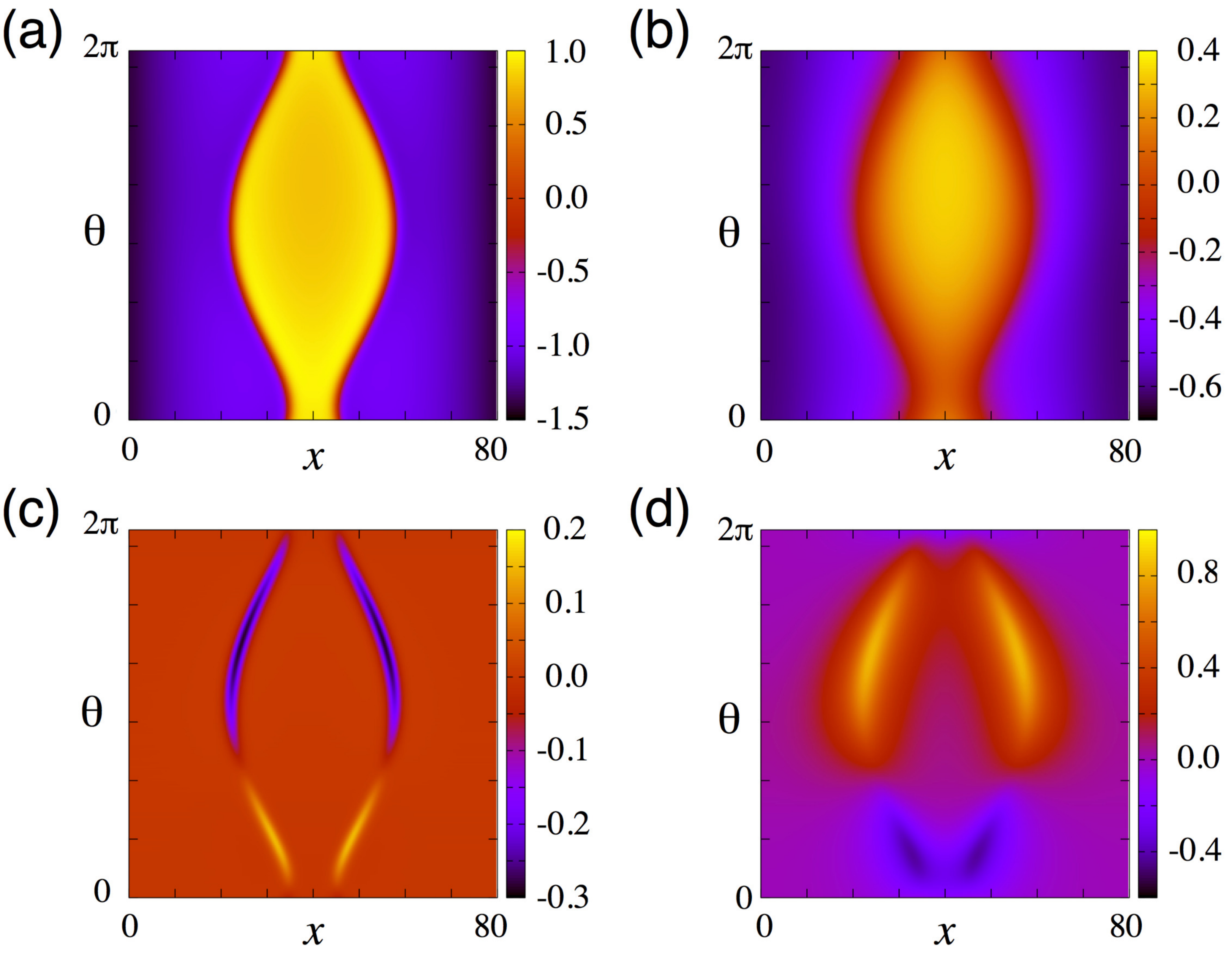}
\caption{Evolution of the limit-cycle solution ${\bm X}_0(x, \theta) = (X_u, X_v)$ and the phase sensitivity function ${\bm Z}(x, \theta) = (Z_u, Z_v)$ for one period of oscillation, $0 \leq \theta \leq 2\pi$.
  (a) $X_u$, (b) $X_v$, (c) $Z_u$, and (d) $Z_v$.}
\label{FigB1}
\end{figure}

\begin{figure}
\centering
\includegraphics[width=0.6\hsize]{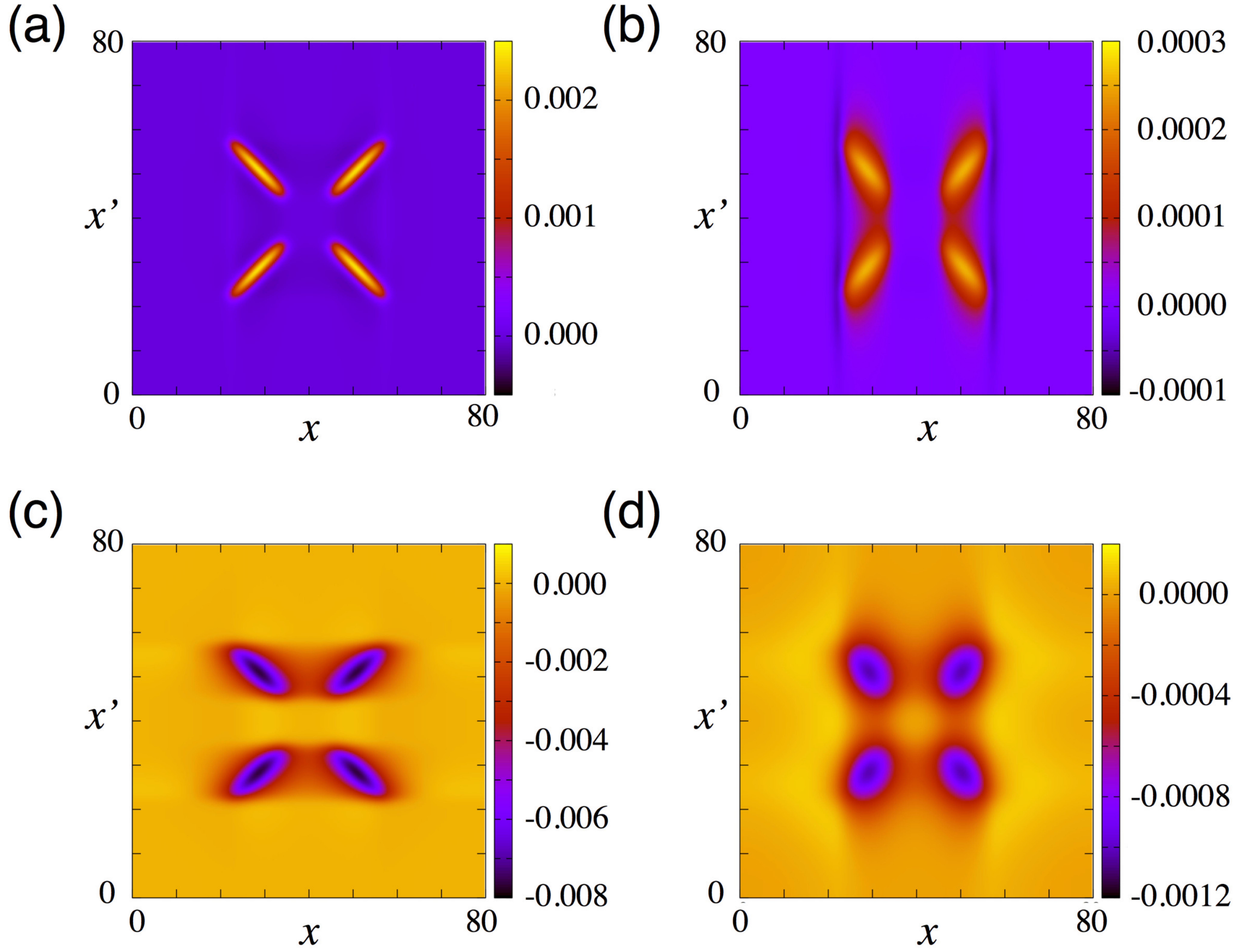}
\caption{Correlation matrix $\hat{W}(x, x') = \left( \begin{array}{cc} W_{uu} & W_{uv} \\ W_{vu} & W_{vv} \end{array} \right)$.
  (a) $W_{uu}$, (b) $W_{uv}$, (c) $W_{vu}$, and (d) $W_{vv}$.}
\label{FigB2}
\end{figure}

\begin{figure}
\centering
\includegraphics[width=0.6\hsize]{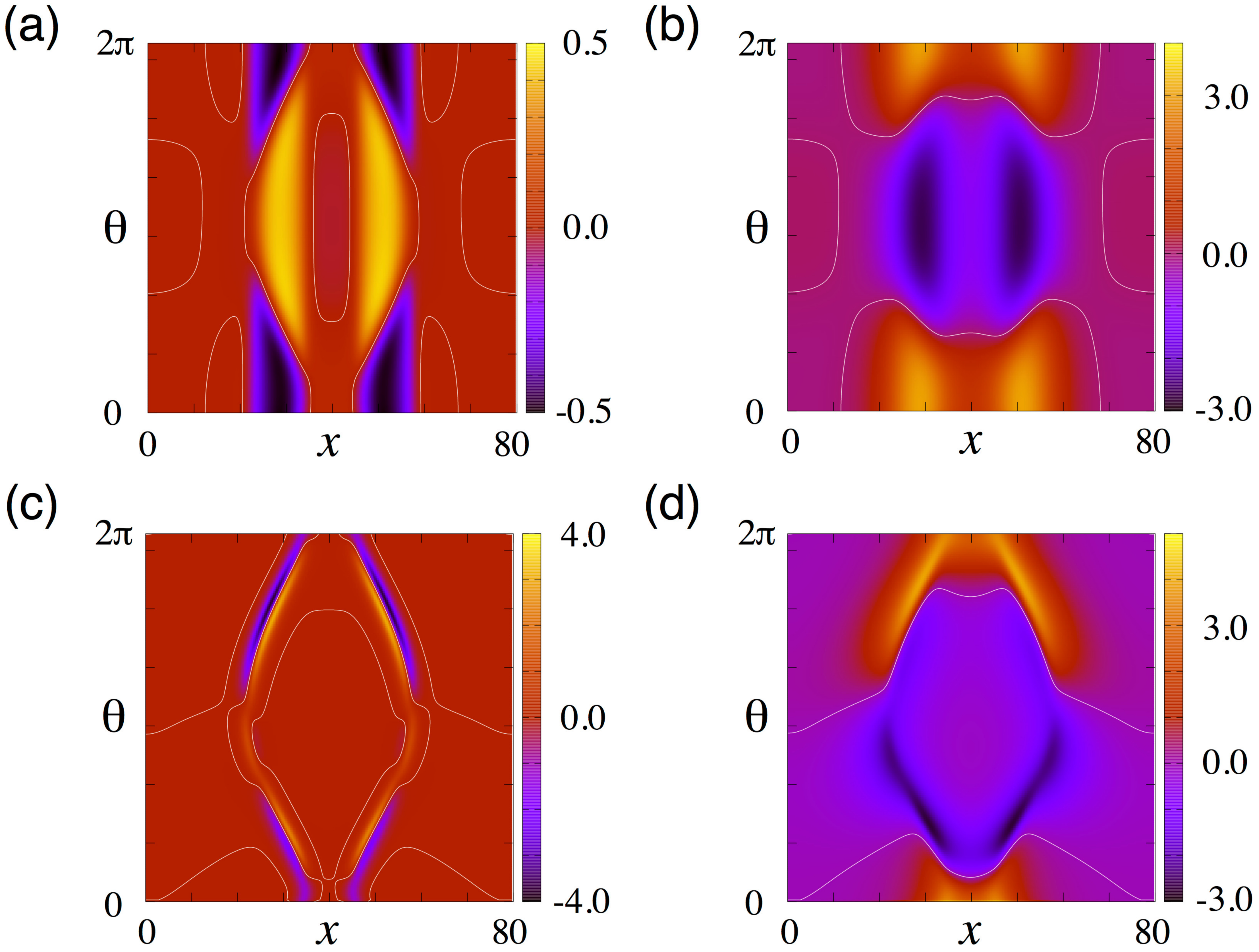}
\caption{Optimized interaction functions of the oscillating spot.
  (a, b) Optimal nonlocal interaction ${\bm G}(x, \theta) = (G_u, G_v)$.
  (a) $G_u$ and (b) $G_v$.
  (c, d) Optimal nonlinear interaction ${\bm H}(x, \theta) = (H_u, H_v)$.
  (c) $H_u$ and (d) $H_v$.
  Evolution for one-period of oscillation are shown ($0 \leq \theta \leq 2\pi$).
  In each figure, the lines indicate the locations where the interaction function vanishes and changes the sign.
  }
\label{FigB3}
\end{figure}

\begin{figure}
\centering
\includegraphics[width=0.55\hsize]{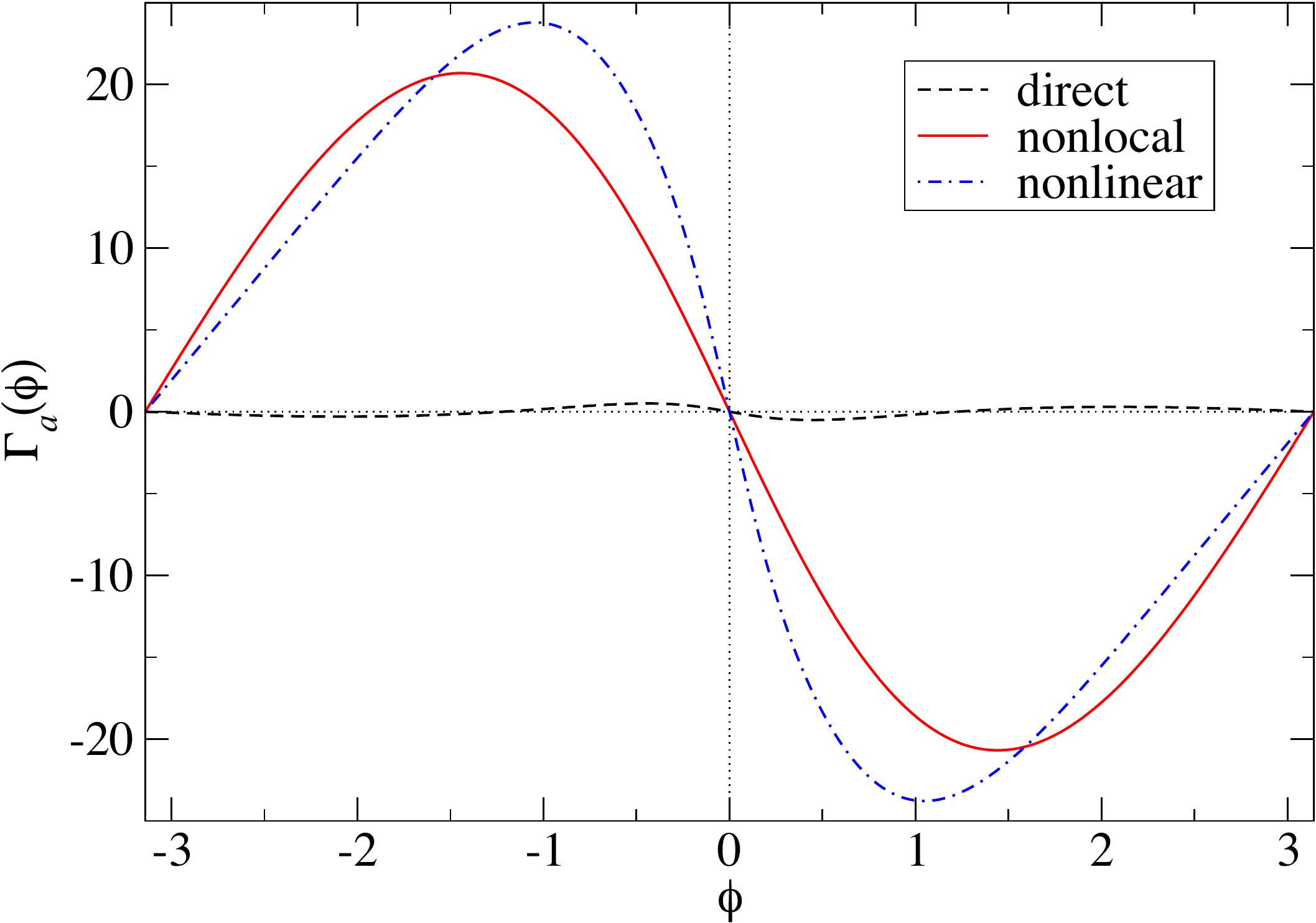}
\caption{Antisymmetric part $\Gamma_a(\phi)$ of the phase coupling functions for the oscillating spot.
  Results for direct interaction, optimal nonlocal interaction, and optimal nonlinear interaction are shown.}
\label{FigB4}
\end{figure}

\begin{figure}
\centering
\includegraphics[width=0.45\hsize]{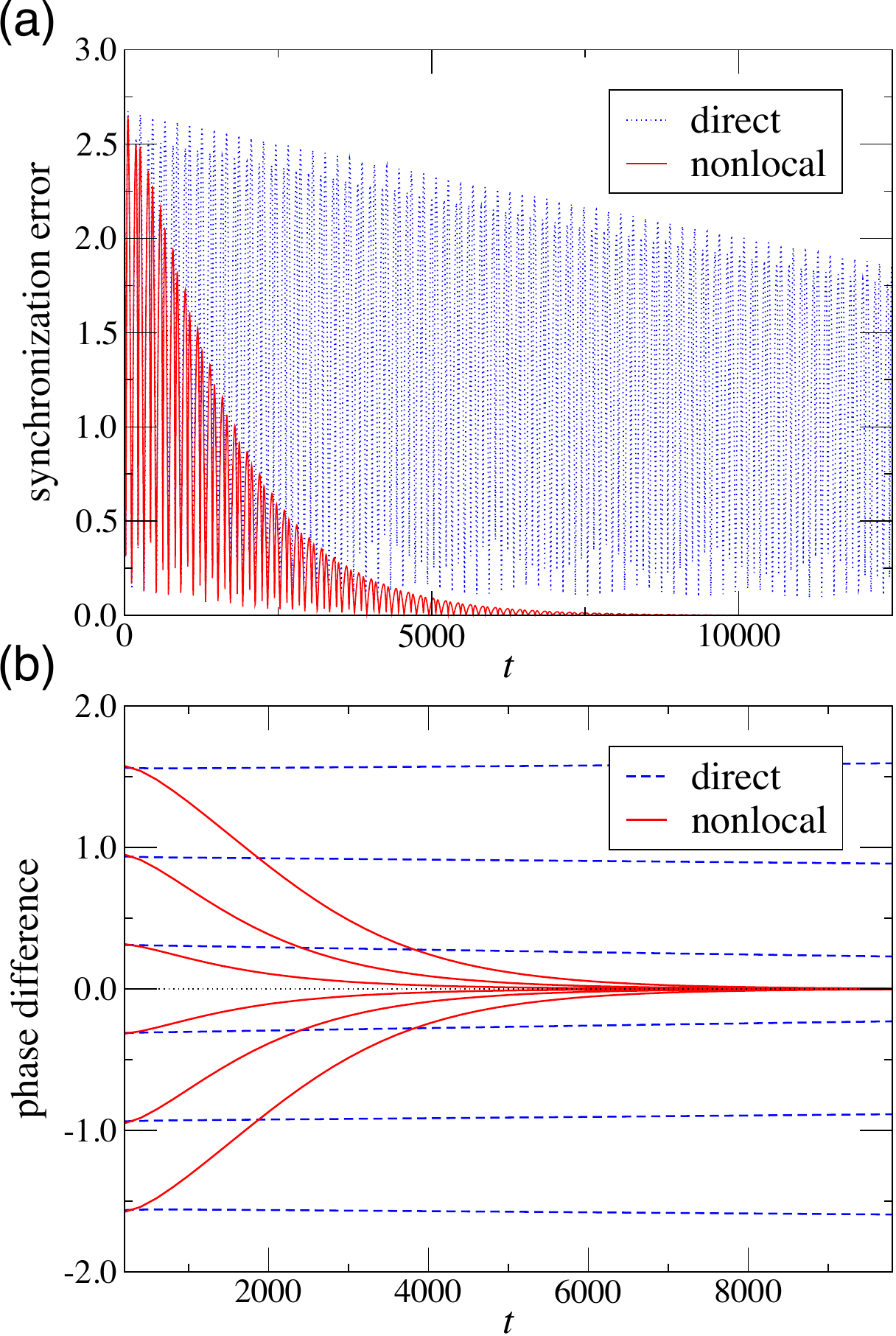}
\caption{Synchronization dynamics between oscillating spots.
  Results for direct interaction and for optimal nonlocal interaction are compared.
  (a) Evolution of synchronization error.
  (b) Evolution of phase difference.}
\label{FigB5}
\end{figure}

Figures~\ref{FigB0}-\ref{FigB5} show the results for the oscillating-spot solution in a one-dimensional space $[0, L]$ with no-flux boundary conditions.
In this case, the parameter $\alpha(x)$ that controls excitability of the media is spatially modulated as $\alpha(x) = \alpha_0 + (\alpha_1 - \alpha_0) (2x/L - 1)^2$ with $\alpha_0 = -1.1$ and $\alpha_1=-1.6$, so that the oscillating spot stays at the center of the system.
The other parameters are $\tau^{-1} = 0.028$, $\gamma = 2.0$, $D_u = 1.0$, and $D_v = 2.5$.
The length of the system is $L=80$ and discretized by using $N=240$ grid points.
The intensity of mutual interaction between the systems is $\epsilon = 0.00001$.

Figure~\ref{FigB0} shows the oscillating-spot solution ${\bm X}_0(x, \theta)$,
the tangent function ${\bm U}(x, \theta) = \partial {\bm X}_0(x, \theta) / \partial \theta$,
and the phase sensitivity function ${\bm Z}(x, \theta)$, all at $\theta = 0$.
In each figure, both $u$ and $v$ components of ${\bm X}_0(x, \theta=0) = (X_u, X_v)$, ${\bm U}(x, \theta=0) = (U_u, U_v)$, and ${\bm Z}(x, \theta=0) = (Z_u, Z_v)$ are plotted. 
The period of the oscillation is $T \approx 200.6$ and the frequency is $\omega \approx 0.0313$.
Figure~\ref{FigB1} shows the evolution of ${\bm X}_0(x, \theta)$ and ${\bm Z}(x, \theta)$ in color code for one period of oscillation, $0 \leq \theta \leq 2\pi$.
The phase sensitivity function is strongly localized near the interfaces of the spot.


Figure~\ref{FigB2} shows all four components of the correlation matrix $\hat{W}(x, x')$, and Fig.~\ref{FigB3} shows the optimal nonlocal interaction ${\bm G}(x, \theta)$ and the optimal nonlinear interaction ${\bm H}(x, \theta)$ for one period of oscillation $(0 \leq \theta \leq 2\pi)$.
The results for the nonlocal and nonlinear cases are different, but the locations at which the interaction functions change their signs are roughly similar in both cases and reflect the locations where the phase sensitivity functions take large values.


Figure~\ref{FigB4} shows the antisymmetric part $\Gamma_a(\phi)$ of the phase coupling function for the direct, optimal nonlocal, and optimal nonlinear interaction.
Both nonlocal interaction and nonlinear interaction yield much higher stability than the direct interaction.
In this case, the optimal nonlocal interaction yields somewhat lower stability than the optimal nonlinear case.
This discrepancy arises because, in addition to the inevitable smoothness of the interaction function due to filtering in the nonlocal case, the spatial linear filter $\hat{A}$ cannot shift the temporal phase of the interaction function ${\bm G}$ from the phase of the oscillation of the spot, while the phase of the optimal nonlinear interaction ${\bm H}$ is slightly shifted from that of the spot.
In this example, while in-phase and anti-phase synchronized solutions are both stable for the direct interaction, both nonlocal interaction and nonlinear interaction give global stability of the in-phase synchronized solution as in the previous case of traveling pulses.
Figure~\ref{FigB5} shows the evolution of the synchronization error and the phase difference between the systems for the direct and optimal nonlocal interactions.
Optimal nonlocal interaction yields much faster convergence to the synchronized state.


\subsection{Rotating spiral}

\begin{figure}
\centering
\includegraphics[width=0.6\hsize]{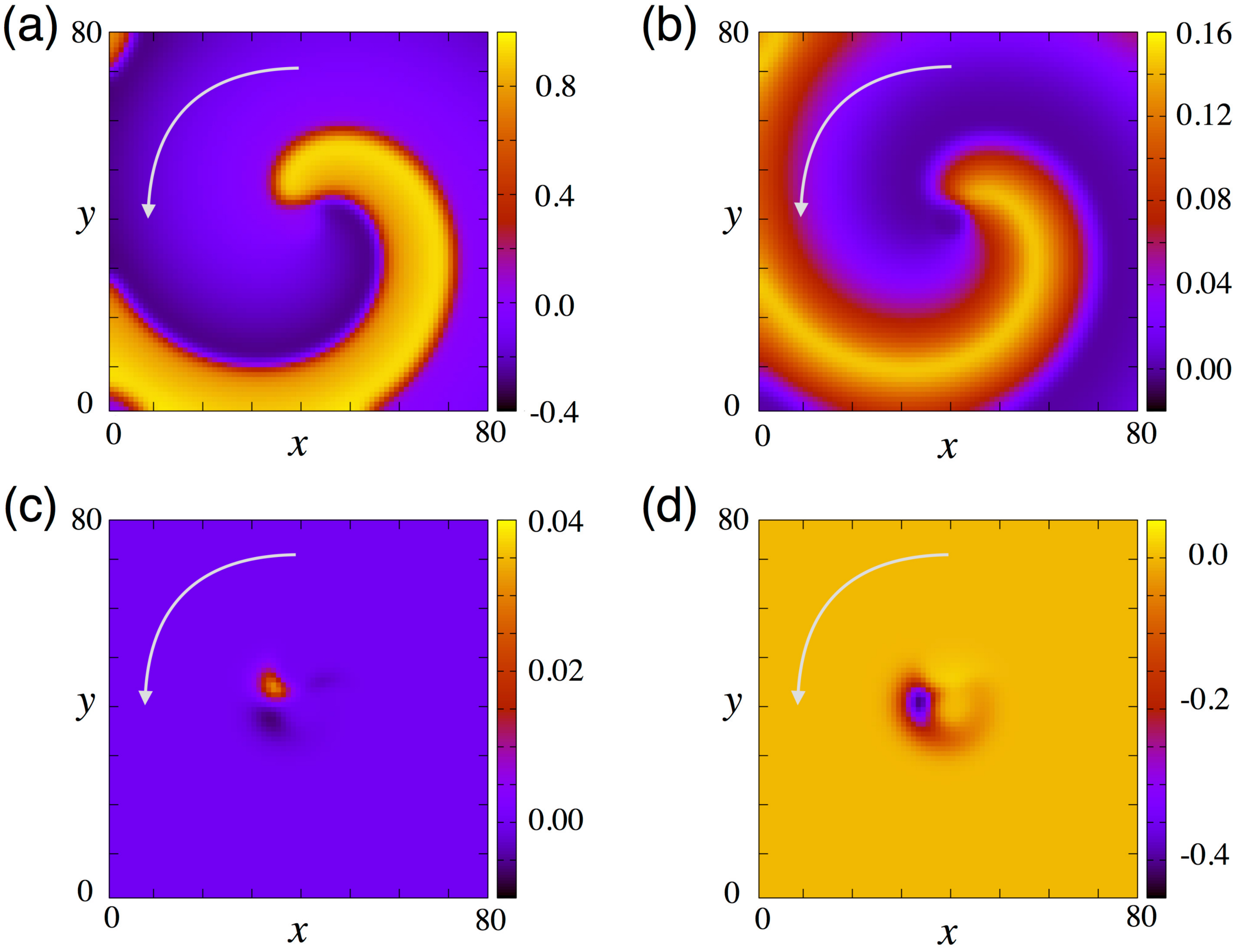}
\caption{Spiral solution of the FHN model.
  (a, b) Snapshots of the limit-cycle solution ${\bm X}_0(x, y, \theta=0) = (X_u, X_v)$.
  (a) $X_u$ and (b) $X_v$.
  (c, d) Snapshots of the phase sensitivity functions ${\bm Z}(x, y, \theta=0) = (Z_u, Z_v)$.
  (c) $Z_u$ and (d) $Z_v$.
  All patterns constantly rotate around the center of the system in the direction shown by the arrow.}
\label{FigC0}
\end{figure}

\begin{figure}
\centering
\includegraphics[width=0.85\hsize]{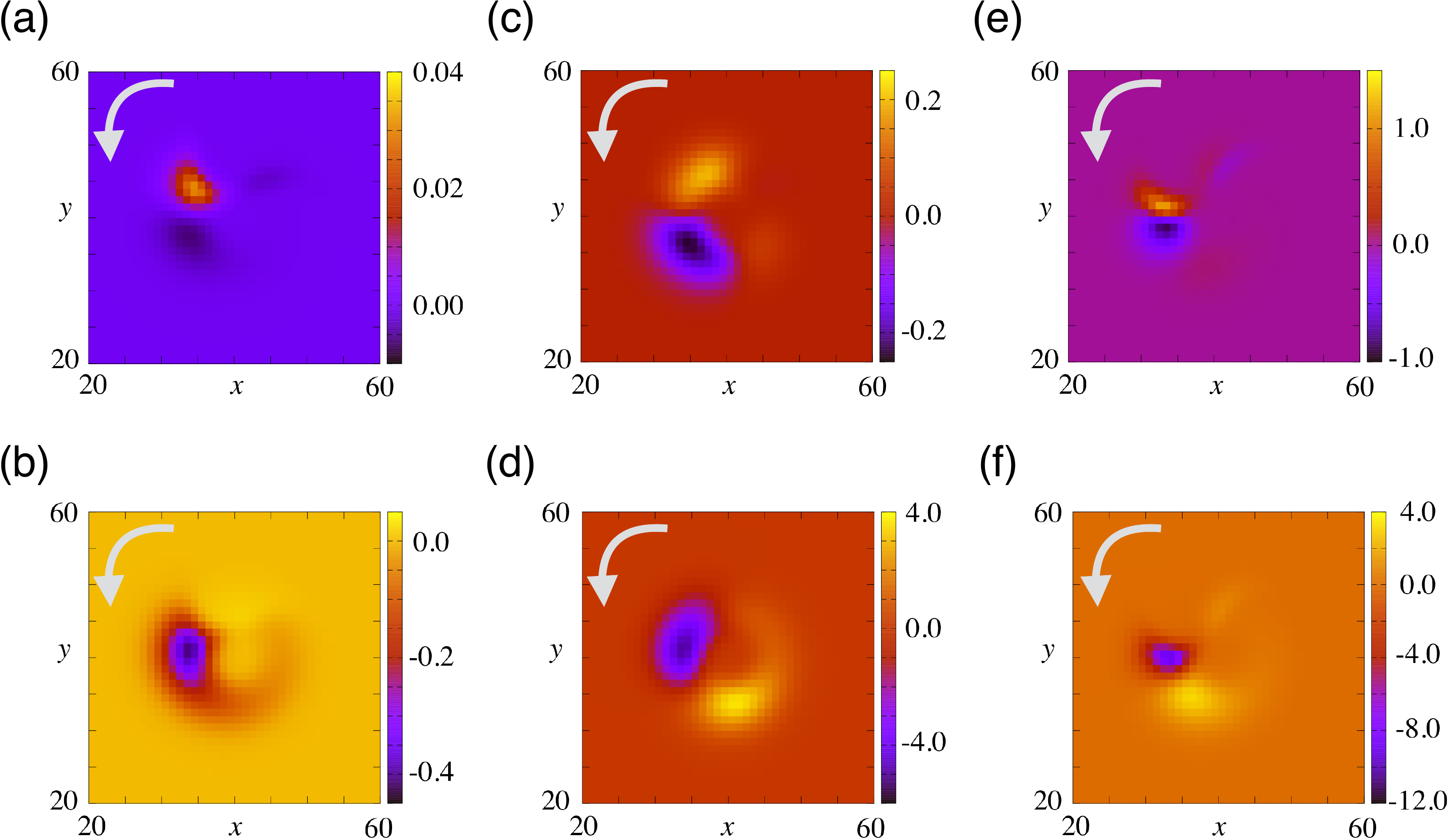}
\caption{
Phase sensitivity function and optimized interaction functions near the spiral tip.
  (a, b) Snapshots of the phase sensitivity function ${\bm Z}(x, y, \theta=0) = (Z_u, Z_v)$.
  (a) $Z_u$ and (b) $Z_v$.
  (c, d) Snapshots of the optimal nonlocal interaction ${\bm G}(x, y, \theta=0) = (G_u, G_v)$.
  (c) $G_u$ and (d) $G_v$.
  (e, f) Snapshots of the optimal nonlinear interaction ${\bm H}(x, y, \theta=0) = (H_u, H_v)$.
  (e) $H_u$ and (f) $H_v$.}
\label{FigC1}
\end{figure}

\begin{figure}
\centering
\includegraphics[width=0.55\hsize]{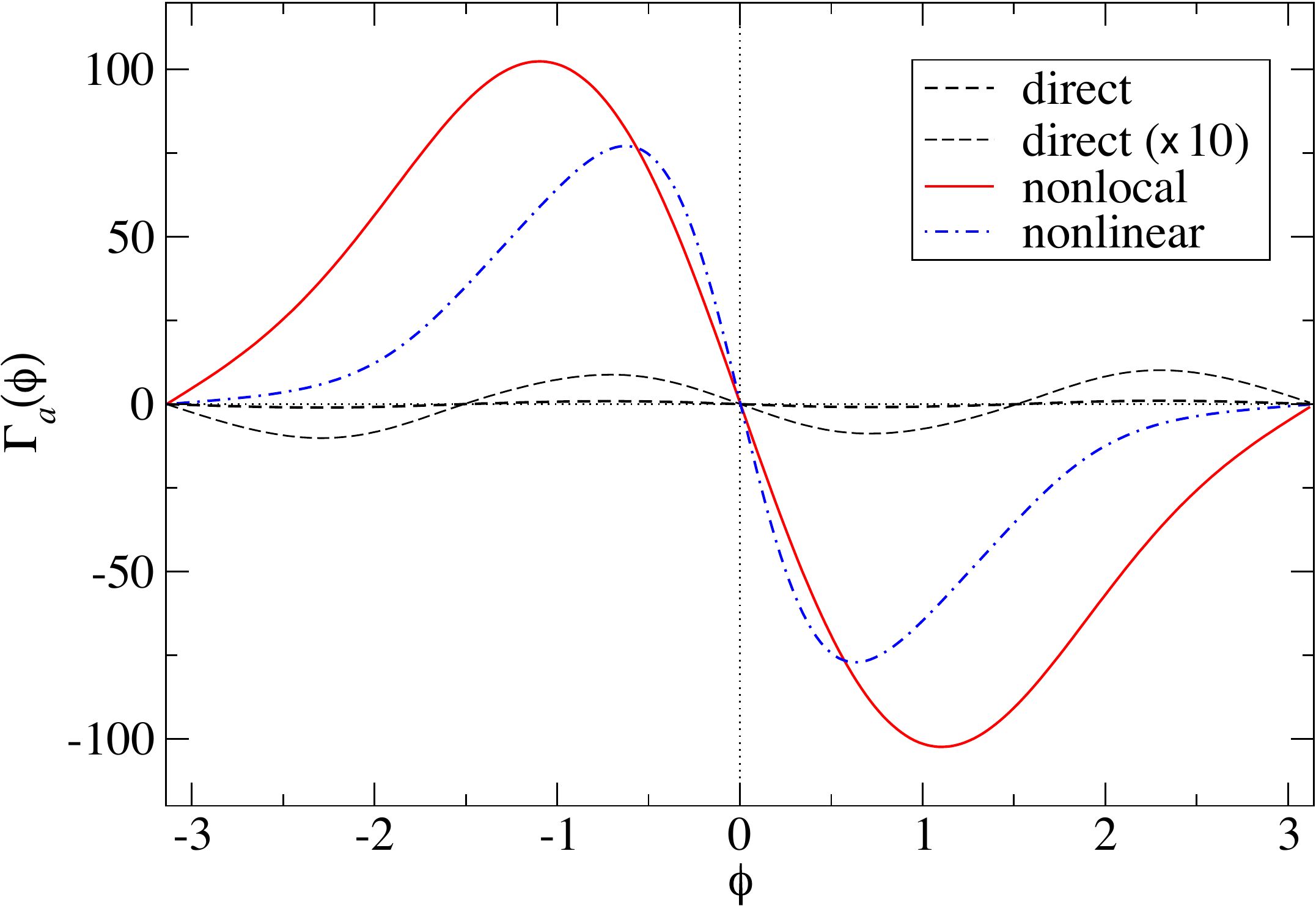}
\caption{Antisymmetric part $\Gamma_a(\phi)$ of the phase coupling functions for the spiral.
  Results for direct interaction, optimal nonlocal interaction, and optimal nonlinear interaction are shown.}
\label{FigC2}
\end{figure}

\begin{figure}
\centering
\includegraphics[width=0.45\hsize]{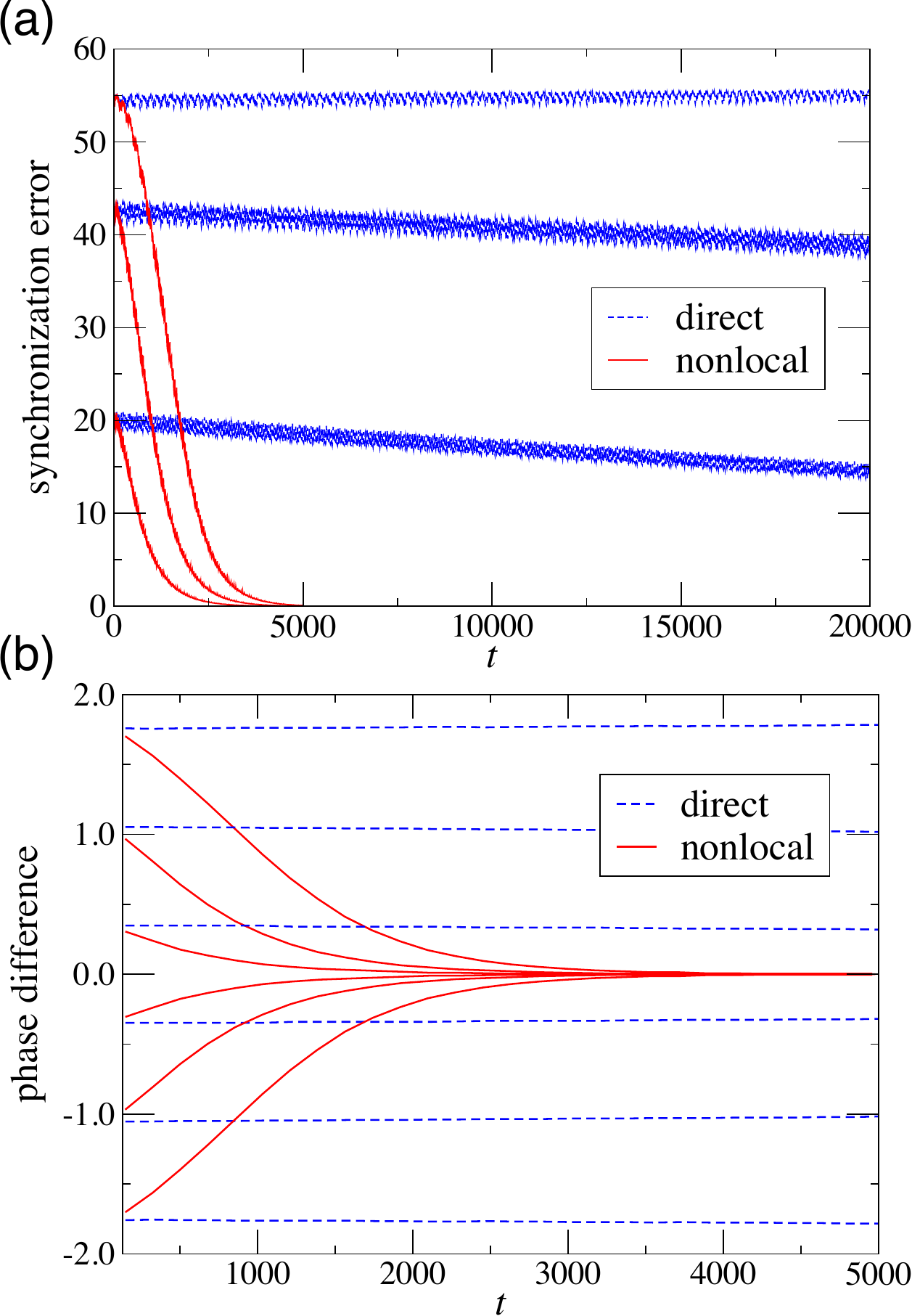}
\caption{Synchronization dynamics between spirals.
  Results for direct interaction and for optimal nonlocal interaction are compared.
  (a) Evolution of synchronization error.
  (b) Evolution of phase difference.}
\label{FigC3}
\end{figure}

This example is motivated by the experimental study by Hildebrand {\it et al.}~\cite{Hildebrand}.
Figures~\ref{FigC0}-\ref{FigC3} show the results for the spiral in a two-dimensional square $0 \leq x, y \leq L$ with no-flux boundary conditions.
The parameter $\alpha(x)$ that controls excitability of the media is spatially modulated as $\alpha(x,y) = \alpha_0 + (\alpha_1 - \alpha_0) \exp(-r^4/r_0^4), r=\sqrt{(x-L/2)^2+(y-L/2)^2}$ with $\alpha_0 = 0.05$ and $\alpha_1=0.5$, so that the spiral stays at the center of the system.
The other parameters are $\tau^{-1} = 0.005$, $\gamma = 2.5$, $D_u = 1.0$, and $D_v = 0.0$.
The size of the system is $L \times L = 80 \times 80$ and discretized by using $N^2=80^2$ grid points.
The interaction intensity between the systems is $\epsilon = 0.00001$.

Figure~\ref{FigC0} shows the spiral solution ${\bm X}_0(x, y, \theta)$ and the phase sensitivity function ${\bm Z}(x, y, \theta)$ at $\theta = 0$.
The spiral pattern ${\bm X}_0(x, y, \theta)$ keeps a constant shape and rotates around the center with a constant frequency and, accordingly, ${\bm U}(x, y, \theta)$ (not shown) and ${\bm Z}(x, y, \theta)$ also rotates around the center.
The period of the oscillation is $T \approx 179$ and the frequency is $\omega \approx 0.0351$.
It can be seen that the phase sensitivity function ${\bm Z}(x, y, \theta)$ is strongly localized near the spiral tip.

Figure~\ref{FigC1} shows the phase sensitivity function ${\bm Z}(x, y, \theta)$, optimal nonlocal interaction function ${\bm G}(x, y, \theta)$, and the optimal nonlinear interaction function ${\bm H}(x, y, \theta)$, enlarged near the spiral tip at $\theta = 0$.
The patterns of ${\bm G}$ and ${\bm H}$ are different, but they are similar in that both of them are localized near the spiral tip and exhibit localized positive and negative spots.


Figure~\ref{FigC2} shows the antisymmetric part $\Gamma_a(\phi)$ of the phase coupling function for the direct, optimal nonlocal, and optimal nonlinear interaction.
Both nonlocal interaction and nonlinear interaction give drastically higher stability than the direct interaction.
The optimal nonlinear interaction yields the highest stability, but the optimal nonlocal interaction also yields reasonably high stability.
In this example, again, the nonlocal interaction and the optimal nonlinear interaction give global stability of the in-phase synchronized solution, while the direct interaction provides bistability of in-phase and anti-phase synchronized solutions.

Finally, Fig.~\ref{FigC3} shows the evolution of the synchronization error and the phase difference between the systems for the direct and optimal nonlocal interactions.
Optimal nonlocal interaction yields much faster convergence to the synchronized state.


\subsection{On global stability of the in-phase synchronized state}

In all considered examples, the optimized interactions yield the in-phase synchronized state as a globally stable fixed point, even if the direct interaction gives bistability of the in-phase and anti-phase states  (see Figs.~\ref{FigA2}, \ref{FigB4}, and \ref{FigC2})~\cite{footnote}.
Since global stability of the in-phase synchronized state is not imposed in the present optimization problem, these results are coincidental, as stated before.
In the present examples, low-order Fourier components are dominant in the waveforms of the oscillation and phase sensitivity function, and thus the phase coupling function $\Gamma(\phi)$ is also a long-wave smooth function of the phase $\phi$.
By maximizing 
$-\Gamma'_a(0)$, the first pair of positive and negative lobes of $\Gamma_a(\phi)$ on both sides of $\phi=0$ axis is further enhanced, resulting in the global stability of the in-phase synchronized state.
In the case of the nonlocal interaction, linear filtering of the field  variable also contributes to the smoothness of the phase coupling function.
In contrast, in the case of the nonlinear interaction,
there is no such smoothing effect by the filtering and the phase coupling function is less smooth than the nonlocal case, as we can observe in Figs.~\ref{FigA2} and \ref{FigC2}.

\section{Summary}

We developed a method for optimizing the interaction function between a pair of mutually interacting reaction-diffusion systems exhibiting stable rhythmic patterns.
We first considered the case with nonlocal interaction and derived the optimal filtering functions for two types of constraints.
We then showed that the optimal nonlinear interaction function is given by the negative of the derivative of the phase sensitivity function in the general case.
Using the FitzHugh-Nagumo reaction-diffusion system, we illustrated that the synchronization between the two systems becomes much faster than the case where all field variables are directly interacting with identical intensity.
These results could be tested experimentally, e.g., by using a pair of spiral patterns in spatially extended photosensitive Belousov-Zhabotinsky reaction~\cite{Hildebrand,ref:showalter15,ref:weiss15,ref:weiss17}.

\begin{acknowledgments}

Y.K. acknowledges financial support from JSPS (Japan) KAKENHI Grant Number JP16K17769.
S.S. acknowledges financial support from JSPS (Japan) KAKENHI Grant Number JP15J12045.
T.Y. acknowledges financial support from JSPS (Japan) KAKENHI Grant Number JP15K05221.
H.N. acknowledges financial support from JSPS (Japan) KAKENHI Grant Numbers JP16H01538, JP16K13847, and JP17H03279.

\end{acknowledgments}

\appendix

\section*{Appendix}

Each component of the optimal nonlocal interaction
\begin{align}
  {\bm G}({\bm r}, t)
  = \frac{1}{\lambda} \int_V d{\bm r}' \, \hat{W}({\bm r}, {\bm r}') {\bm X}({\bm r}', t)
  \label{linear1}
\end{align}
can be expressed as
\begin{align}
  G_i({\bm r}, t)
  &= \frac{1}{\lambda} \int_V d{\bm r}' \sum_{j=1}^m W_{ij}({\bm r}, {\bm r}') X_j({\bm r}', t)
  \cr
  &= \frac{1}{2\pi\lambda} \int_0^{2\pi} d\psi \int_V d{\bm r}' \sum_{j=1}^m
  Z_i({\bm r}, \psi) U_j({\bm r}', \psi) X_j({\bm r}', t)
  \cr
  &= \frac{1}{2\pi\lambda} \int_0^{2\pi} d\psi \, Z_i({\bm r}, \psi) \bar{X}(\psi, t),
\end{align}
where
\begin{align}
  \bar{X}(\psi, t)
  = \int_V d{\bm r}' \sum_{j=1}^m U_j({\bm r}', \psi) X_j({\bm r}', t).
\end{align}
That is, the optimal nonlocal interaction can also be represented as
\begin{align}
  {\bm G}({\bm r}, t)
  = \frac{1}{2\pi\lambda} \int_0^{2\pi} d\psi \, {\bm Z}({\bm r}, \psi) \bar{X}(\psi, t)
  \label{linear2}
\end{align}
and we use this expression in the numerical simulations.
When we discretize a $d$-dimensional spatial domain by using $N^d$ grid points and one period of oscillation by using $M$ points, the computational costs of Eq.~(\ref{linear1}) and Eq.~(\ref{linear2}) are $O(N^{2d})$ and $O(M N^d)$, respectively.
Therefore, Eq.~(\ref{linear2}) is much more efficient than Eq.~(\ref{linear1}) when the spatial dimension $d$ of the system is large.


\begin{thebibliography}{99}

\bibitem{ref:winfree80}
  A.~T.~Winfree,
  {\it The Geometry of Biological Time}
  (Springer, New York, 1980; Springer, Second Edition, New York, 2001).

\bibitem{Kuramoto}
  Y.~Kuramoto,
  {\it Chemical Oscillations, Waves, and Turbulence}
  (Springer, New York, 1984; Dover, New York, 2003).

\bibitem{Glass}
L. Glass and M. C. Mackey,
{\it From clocks to chaos: the rhythms of life}
(Princeton University Press, Princeton, 1988).

\bibitem{ref:pikovsky01}
  A.~Pikovsky, M.~Rosenblum, and J.~Kurths,
  {\it Synchronization: A Universal Concept in Nonlinear Sciences}
  (Cambridge University Press, Cambridge, 2001).

\bibitem{ref:hoppensteadt97}
  F.~C.~Hoppensteadt and E.~M.~Izhikevich,
  {\it Weakly Connected Neural Networks}
  (Springer, New York, 1997).
  
\bibitem{ref:ermentrout10}
  G.~B.~Ermentrout and D.~H.~Terman,
  {\it Mathematical Foundations of Neuroscience}
  (Springer, New York, 2010).

\bibitem{bullo}
 F. D\"orfler and F. Bullo,
 Synchronization and transient stability in power networks and nonuniform Kuramoto oscillators,
 SIAM Journal on Control and Optimization {\bf 50}, 1616-1642 (2012).

\bibitem{Kotani}
K. Kotani {\it et al.},
Adjoint Method Provides Phase Response Functions for Delay-Induced Oscillations,
Phys. Rev. Lett. {\bf 109}, 044101 (2012).

\bibitem{Novicenko}
V. Novi{\v c}enko and K. Pyragas,
Phase reduction of weakly perturbed limit cycle oscillations in time-delay systems,
Physica D {\bf 241}, 1090–1098 (2012).

\bibitem{Nakao}
  H.~Nakao, T.~Yanagita, and Y.~Kawamura,
  Phase-reduction approach to synchronization of spatiotemporal rhythms in reaction-diffusion systems,
  Phys. Rev. X {\bf 4}, 021032 (2014).

\bibitem{Fukushima}
	S. Fukushima, S. Nakanishi, K. Fukami, S. I. Sakai, T. Nagai, T. Tada, and Y. Nakato,
	Observation of synchronized spatiotemporal reaction waves in coupled 	electrochemical oscillations of an NDR type,
	Electrochem. Comm. {\bf 7}, 411 (2005). 

\bibitem{Hildebrand}
  M.~Hildebrand, J.~Cui, E.~Mihaliuk, J.~Wang, and K.~Showalter,
  Synchronization of spatiotemporal patterns in locally coupled excitable media,
  Phys. Rev. E {\bf 68}, 026205 (2003).

  \bibitem{Kawamura2} 
  Y.~Kawamura and H.~Nakao,
  Collective phase description of oscillatory convection,
  Chaos {\bf 23}, 043129 (2013).
  Y.~Kawamura and H.~Nakao,
  Phase description of oscillatory convection with a spatially translational mode,
  Physica D {\bf 295-296}, 11 (2015).
  
  \bibitem{read2010}
	A.~A.~Castrejo\'on-Pita and P.~L.~Read,
	Synchronization in a pair of thermally coupled rotating baroclinic annuli:
	understanding atmospheric teleconnections in the laboratory,
	Phys. Rev. Lett. {\bf 104}, 204501 (2010).

\bibitem{read2012}
	P.~L.~Read and A.~A.~Castrejo\'on-Pita,
	Phase synchronization between stratospheric and tropospheric quasi-biennial and semi-annual oscillations,
	Q. J. R. Meteorol. Soc. {\bf 138}, 1338-1349 (2012).

\bibitem{ref:brown04}
  E.~Brown, J.~Moehlis, and P.~Holmes,
  On the phase reduction and response dynamics of neural oscillator populations,
  Neural Comput. {\bf 16}, 673 (2004).
  
\bibitem{ref:nakao16}
  H.~Nakao,
  Phase reduction approach to synchronization of nonlinear oscillators,
  Contemp. Phys. {\bf 57}, 188 (2016).
  
\bibitem{ref:ashwin16}
  P.~Ashwin, S.~Coombes, and R.~Nicks,
  Mathematical frameworks for oscillatory network dynamics in neuroscience,
  J. Math. Neurosci. {\bf 6}, 1 (2016).
  
\bibitem{ref:mikhailov06}
  A.~S.~Mikhailov and K.~Showalter,
  Control of waves, patterns and turbulence in chemical systems,
  Phys. Rep. {\bf 425}, 79 (2006).
  
\bibitem{ref:mikhailov13}
  A.~S.~Mikhailov and G.~Ertl (Editors),
  {\it Engineering of Chemical Complexity}
  (World Scientific, Singapore, 2013).
  
\bibitem{ref:mikhailov14}
  A.~S.~Mikhailov and G.~Ertl (Editors),
  {\it Engineering of Chemical Complexity II}
  (World Scientific, Singapore, 2014).
  
\bibitem{ref:cross93}
  M.~C.~Cross and P.~C.~Hohenberg,
  Pattern formation outside of equilibrium,
  Rev. Mod. Phys. {\bf 65}, 851 (1993).
  
\bibitem{ref:cross09}
  M.~C.~Cross and H.~Greenside,
  {\it Pattern Formation and Dynamics in Nonequilibrium Systems}
  (Cambridge University Press, Cambridge, 2009).

\bibitem{ref:engel14}
 J. L\"ober, R. Coles, J. Siebert, H. Engel, and E. Sch\"oll,
 Control of Chemical Wave Propagation,
 in {\it Engineering of Chemical Complexity II}, pp. pp. 185--207, World Scientific, 2014.

\bibitem{ref:engel16}
	C. Ryll, J. L\"ober, S. Martens, H. Engel, and F. Tr\"oltzsch,
	Analytical, Optimal, and Sparse Optimal Control of Traveling Wave Solutions to Reaction-Diffusion Systems,
	in {\it Control of Self-Organizing Nonlinear Systems}, pp. 189--210, Springer, 2016.
  
\bibitem{Sakurai}
T. Sakurai, E. Mihaliuk, F. Chirila, K. Showalter,
Design and control of wave propagation patterns in excitable media,
Science {\bf 296}, 2009-2012 (2002).

\bibitem{Epstein}
	I.~R.~Epstein, I.~B.~Berenstein, M.~Dolnik, V.~K.~Vanag, L.~Yang, A.~M.~ 	Zhabotinsky,
	Coupled and forced patterns in reaction-diffusion systems,
	Phil. Trans. R. Soc. A {\bf 366}, 397–408 (2008).

\bibitem{ref:showalter15}
  K.~Showalter and I.~R.~Epstein,
  From chemical systems to systems chemistry: Patterns in space and time,
  Chaos {\bf 25}, 097613 (2015).

\bibitem{ref:weiss15}
  S.~Weiss and R.~D.~Deegan,
  Quantized orbits in weakly coupled Belousov-Zhabotinsky reactors,
  Europhys. Lett. {\bf 110}, 60004 (2015).
  
\bibitem{ref:weiss17}
  S.~Weiss and R.~D.~Deegan,
  Weakly and strongly coupled Belousov-Zhabotinsky patterns,
  Phys. Rev. E {\bf 95}, 022215 (2017).

\bibitem{Kawamura1}
  Y.~Kawamura, H.~Nakao, K.~Arai, H.~Kori, and Y.~Kuramoto,
  Collective phase sensitivity,
  Phys. Rev. Lett. {\bf 101}, 024101 (2008).
  Y.~Kawamura, H.~Nakao, K.~Arai, H.~Kori, and Y.~Kuramoto,
  Phase synchronization between collective rhythms of globally coupled oscillator groups: Noisy identical case,
  Chaos {\bf 20}, 043109 (2010).
  Y.~Kawamura, H.~Nakao, and Y.~Kuramoto,
  Collective phase description of globally coupled excitable elements,
  Phys. Rev. E {\bf 84}, 046211 (2011).
  Y.~Kawamura,
  Collective phase dynamics of globally coupled oscillators: Noise-induced anti-phase synchronization,
  Physica D {\bf 270}, 20 (2014).
  Y.~Kawamura,
  Collective phase reduction of globally coupled noisy dynamical elements,
  Phys. Rev. E {\bf 95}, 032225 (2017).
  
\bibitem{Kawamura3}
  Y.~Kawamura and H.~Nakao,
  Noise-induced synchronization of oscillatory convection and its optimization,
  Phys. Rev. E {\bf 89}, 012912 (2014).
  
\bibitem{Zlotnik}
  A.~Zlotnik, Y.~Chen, I.~Z.~Kiss, H.~Tanaka, and J.-S.~Li,
  Optimal waveform for fast entrainment of weakly forced nonlinear oscillators,
  Phys. Rev. Lett. {\bf 111}, 024102 (2013).

\bibitem{ref:moehlis06}
  J.~Moehlis, E.~Shea-Brown, and H.~Rabitz,
  Optimal inputs for phase models of spiking neurons,
  J. Comput. Nonlin. Dyn. {\bf 1}, 358 (2006).
  
\bibitem{ref:harada10}
  T.~Harada, H.-A.~Tanaka, M.~J.~Hankins, and I.~Z.~Kiss,
  Optimal waveform for the entrainment of a weakly forced oscillator,
  Phys. Rev. Lett. {\bf 105}, 088301 (2010).
  
\bibitem{ref:dasanayake11}
  I.~Dasanayake and J.-S.~Li,
  Optimal design of minimum-power stimuli for phase models of neuron oscillators,
  Phys. Rev. E {\bf 83}, 061916 (2011).
  
\bibitem{ref:zlotnik12}
  A.~Zlotnik and J.-S.~Li,
  Optimal entrainment of neural oscillator ensembles,
  J. Neural Eng. {\bf 9}, 046015 (2012).
  
\bibitem{ref:pikovsky15-prl}
  A.~Pikovsky,
  Maximizing coherence of oscillations by external locking,
  Phys. Rev. Lett. {\bf 115}, 070602 (2015).
         
\bibitem{ref:tanaka14a}
  H.-A.~Tanaka,
  Synchronization limit of weakly forced nonlinear oscillators,
  J. Phys. A: Math. Theor. {\bf 47}, 402002 (2014).
  
\bibitem{ref:tanaka14b}
  H.-A.~Tanaka,
  Optimal entrainment with smooth, pulse, and square signals in weakly forced nonlinear oscillators,
  Physica D {\bf 288}, 1 (2014).
  
\bibitem{ref:tanaka15}
  H.-A.~Tanaka, I.~Nishikawa, J.~Kurths, Y.~Chen, and I.~Z.~Kiss,
  Optimal synchronization of oscillatory chemical reactions
  with complex pulse, square, and smooth waveforms signals maximizes Tsallis entropy,
  Europhys. Lett. {\bf 111}, 50007 (2015).
  
\bibitem{ref:zlotnik2016}
  A.~Zlotnik, R.~Nagao, I.~Z.~Kiss, and J.-S.~Li,
  Phase-selective entrainment of nonlinear oscillator ensembles,
  Nature Communications {\bf 7}, 10788 (2016).
  
\bibitem{Shirasaka}
  S.~Shirasaka, N.~Watanabe, Y.~Kawamura, and H.~Nakao,
  Improving stability of mutual synchronization in weakly coupled limit-cycle oscillators,
  Phys. Rev. E (2017) (to be published).


\bibitem{footnote}
The bistability can be interpreted from the functional shapes of the phase sensitivity functions and the patterns. In the case of traveling pulses, the phase sensitivity $Z_u$ takes positive values only in a small region in front of the pulse. Thus, when the two pulses are roughly aligned and one of them is slightly ahead of the other, the effect from the faster pulse ($X_u > 0$) pushes the slower pulse forward and they tend do align again. When the two pulses are sufficiently separated, they tend to repel each other, because the effect from the other pulse is negative ($X_u < 0$) and the faster pulse pushes the slower pulse backward. Because of the periodic boundary conditions, the two pulses settle in the completely anti-phase state. Similar mechanisms also work for the rotating spirals and oscillating spots.
	  
\end{thebibliography}
\end{document}